\documentclass[
aps,
prl,
reprint,
superscriptaddress,
longbibliography
]{revtex4-2}
\usepackage{color}
\usepackage[dvipsnames]{xcolor}

\usepackage{graphicx}
\graphicspath{{images/}}

\usepackage{comment}

\usepackage[T1]{fontenc}
\usepackage[utf8]{inputenc}

\usepackage{amsmath,amssymb,amsfonts}

\usepackage{newtxtext}
\usepackage{newtxmath}

\usepackage{bm}
\usepackage{braket}
\usepackage{physics}
\usepackage{microtype}

\usepackage{hyperref}
\hypersetup{
    colorlinks=true,
    linkcolor=MidnightBlue,
    citecolor=MidnightBlue,
    urlcolor=MidnightBlue
}

\begin{document} 

    \title{Ultra-Weak-Pump Microwave-to-Optical Quantum Transduction via a Single Color Center}
        
        \author{Kyosuke Goto}
        \affiliation{Department of Physics, Graduate School of Engineering Science, Yokohama National University, 79-5 Tokiwadai, Hodogaya, Yokohama, Kanagawa, 240-8501, Japan}
        
        \author{Hodaka Kurokawa}
        \affiliation{Quantum Information Research Center, Institute of Advanced Sciences, Yokohama National University, 79-5 Tokiwadai, Hodogaya, Yokohama, Kanagawa, 240-8501, Japan}
        \affiliation{Institute of Industrial Science (IIS), The University of Tokyo, 4-6-1 Komaba, Meguro-ku, Tokyo 153-8505, Japan}

        \author{Hideo Kosaka}
        \email{kosaka-hideo-yp@ynu.ac.jp}
        \affiliation{Department of Physics, Graduate School of Engineering Science, Yokohama National University, 79-5 Tokiwadai, Hodogaya, Yokohama, Kanagawa, 240-8501, Japan}
        \affiliation{Quantum Information Research Center, Institute of Advanced Sciences, Yokohama National University, 79-5 Tokiwadai, Hodogaya, Yokohama, Kanagawa, 240-8501, Japan}

        \author{Kazuki Koshino}
        \email{kazuki.koshino@opticamember.org}
        \altaffiliation[Present address: ]{Fujitsu Research Laboratories, Fujitsu Limited, Atsugi 243-0197, Japan}
        \affiliation{Institute for Liberal Arts, Institute of Science Tokyo, 2-8-30 Konodai, Ichikawa, Chiba 272-0827, Japan}

\begin{abstract}

    Scaling up superconducting quantum processors remains a central challenge for realizing fault-tolerant quantum computation.
    Although distributed architectures based on optical photons offer a promising route to scalability, they require an efficient microwave-to-optical quantum transducer that operates at cryogenic temperatures.
    Existing approaches typically rely on strong optical pumping, which induces undesirable heating and degrades single-photon coherence.
    Here, we propose a microwave-to-optical quantum transducer based on resonant scattering at a single color center embedded in a diamond optomechanical resonator.
    We show that strong coupling between the color center and the optical cavity enables coherent conversion at extremely weak pump powers on the order of 10 pW.
    The proposed scheme enables remote entanglement generation at a kilohertz-scale rate with a fidelity exceeding 0.9, demonstrating a viable pathway toward ultra-weak-pump and high-efficiency quantum transducers based on a solid-state defect.

\end{abstract}

    \maketitle

    Fault-tolerant quantum computing is expected to require at least several hundred thousand physical qubits \cite{Gidney_Quantum2021}, posing a major challenge for modern quantum science and technology.
    To address this scalability issue, distributed quantum computing architectures \cite{Nickerson_NatCommun2013, Main_Nature2025, Weaver_arXiv2025} have attracted increasing attention across several platforms.
    For superconducting processors operated at millikelvin temperatures using microwave tones, optical interconnects are attractive because they could enable quantum state transfer over flexible and broadband optical fibers at room temperature.
    Realizing such interconnects, however, requires microwave-to-optical quantum transducers that simultaneously achieve high conversion efficiency and low added noise, which remains a central challenge.
 
    In recent years, substantial progress has been made in quantum transducers based on electro-optics \cite{Xu_NatCommun2021, Sahu_NatCommun2022, Warner_NatPhys2025}, optomechanics \cite{Higginbotham_NatPhys2018, Mirhosseini_Nature2020, Jiang_NatPhys2023, Meesala_NatPhys2024, Weaver_Natnano2024, Thiel_NatPhys2025, Zhao_Natnano2025}, magnonics \cite{Hisatomi_PRB2016, Ihn_PRB2020, Zhu_Optica2020}, atomic ensembles \cite{Han_PRL2018, Vogt_PRA2019, Tu_NatPhotonics2022, Kumar_NatPhys2023, Borowka_NatPhotonics2024}, and spin ensembles \cite{Rochman_NatCommun2023, Xie_NatPhys2025, Khalifa_npj2025}.
    In resonator-based approaches, a common strategy is to employ small mode volumes $V$ to enhance the vacuum coupling $g$ between the target modes, while using high-$Q$ resonators to suppress unwanted single-photon loss.
    Increasing $g$ reduces the pump power required for efficient conversion and thereby mitigates pump-induced added noise.
    Nevertheless, suppressing the added noise below one photon generally requires operating at weak pump powers, at the expense of conversion efficiency \cite{Sahu_NatCommun2022, Zhao_Natnano2025}.
    Resolving this efficiency-noise trade-off is thus essential for practical microwave-to-optical quantum transduction \cite{Arnold_NatCommun2020, Stockill_NatCommun2022, Burger_APL2025, Khalifa_npj2025, Li_arXiv2025}.
 
\begin{figure}[htbp]
        \begin{center}
        \includegraphics[width=\columnwidth, trim=15mm 17mm 15mm 13mm, clip]{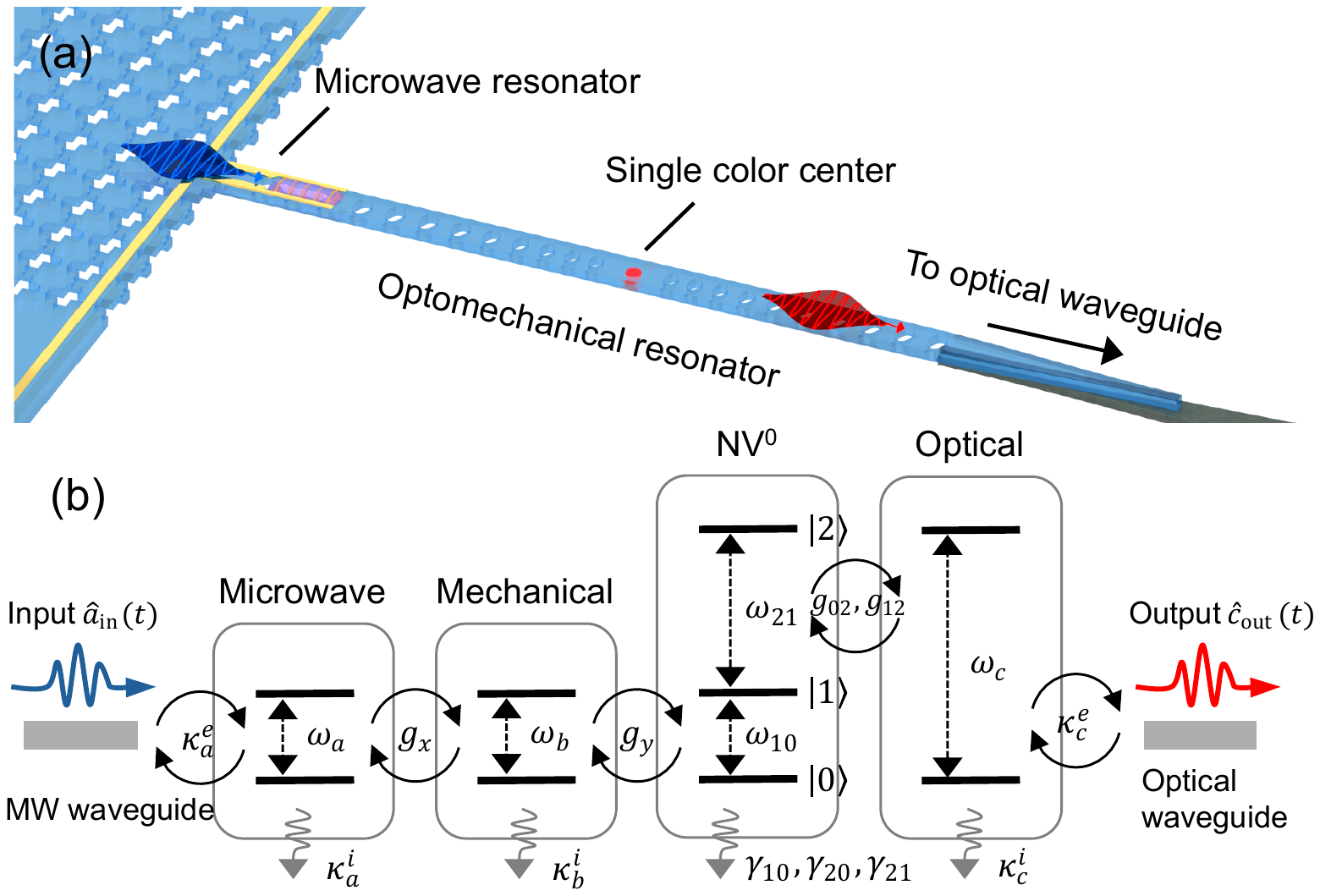}
        \caption{(a) Schematic of the proposed device.
                 (b) Mode-coupling diagram under the triply resonant condition $\omega_a = \omega_b = \omega_{10}$ and $\omega_c = \omega_{20}$;
                 the pump tone is tuned to $\omega_{21}$.
                } \label{FIG:schematic}
        \end{center}
    \end{figure}

    Here we propose an extremely weak-pump quantum transducer based on resonant scattering at a single neutral nitrogen-vacancy (NV$^0$) center in a diamond optomechanical resonator [Fig.~\ref*{FIG:schematic}(a)].
    This scheme exploits the strong coupling of NV$^0$ to both optical and strain fields by combining the small mode volume of an optomechanical resonator with the resonant dipole transitions of NV$^0$.
    The resulting vacuum coupling strength is more than three orders of magnitude larger than that in conventional optomechanical systems, enabling highly efficient conversion under weak pumping.
    Steady-state analysis presented in this paper shows a coherent conversion efficiency of $0.32$ at a pump power of $48$~pW.

    We further analyze remote entanglement generation, a pivotal resource for distributed quantum computing \cite{Kurpiers_Nature2018}, by modeling each transducer as a quantum channel within a single-click scheme \cite{Cabrillo_PRA1999}.
    Our analysis predicts a fidelity of $0.91$ and a heralding rate of $5.3$~kHz at a repetition rate of $1$~MHz, highlighting the potential of the proposed transducer for distributed superconducting quantum computing.

 	The proposed quantum transducer consists of a microwave resonator $\hat{a}$ (frequency $\omega_a/2\pi$), a mechanical resonator $\hat{b}$ (frequency $\omega_b/2\pi$), an NV$^0$ center, and an optical resonator $\hat{c}$ (frequency $\omega_c/2\pi$), coupled in series as shown in Fig.~\ref*{FIG:schematic}(b).
    The NV$^0$ center has four ground states formed by orbital states ($|+\rangle_{\mathrm{o}}, |-\rangle_{\mathrm{o}}$) and spin states ($\lvert\uparrow\rangle_{\mathrm{s}},\lvert\downarrow\rangle_{\mathrm{s}}$) \cite{Baier2020,Kurokawa2024,Kurokawa2025}.
    We define $\hat{L}_{z} = |+\rangle_{\mathrm{o}}{}_{\mathrm{o}}\langle+| - |-\rangle_{\mathrm{o}}{}_{\mathrm{o}}\langle-|$ and $\hat{L}_{\pm} = |\pm\rangle_{\mathrm{o}}{}_{\mathrm{o}}\langle \mp|$ in the orbital subspace, and $\hat{S}_{z} = (\lvert\uparrow\rangle_{\mathrm{s}}{}_{\mathrm{s}}\langle\uparrow\rvert - \lvert\downarrow\rangle_{\mathrm{s}}{}_{\mathrm{s}}\langle\downarrow\rvert)/2$ in the spin subspace, respectively.
    The zero-field Hamiltonian is then given by $\hat{H}_{\mathrm{NV^0}}/\hbar = 2\lambda_{\mathrm{so}}\hat{L}_{z}\hat{S}_{z} + \varepsilon_{\perp}(\hat{L}_{+}+\hat{L}_{-})$, where the two terms describe spin-orbit coupling and static strain, respectively.
    Under a finite transverse strain, the ground states split into two spin-degenerate branches separated by $\Delta_{\mathrm{gs}} = 2\sqrt{\lambda_{\mathrm{so}}^2 + \varepsilon_{\perp}^2}$.
    Neglecting spin, NV$^0$ reduces to a three-level system with lower and upper branch manifolds $|0\rangle$ and $|1\rangle$, and the orbital excited-state manifold $|2\rangle$.
   
    The system Hamiltonian is then written as
    \begin{equation}
        \label{eq:system_Hamiltonian}
        \begin{split}
            \hat{H}/\hbar =&\,\omega_{a} \hat{a}^{\dagger} \hat{a} +
                               \omega_{b} \hat{b}^{\dagger} \hat{b} + 
                               \omega_{c} \hat{c}^{\dagger} \hat{c} +
                               \omega_{10}\hat{\sigma}_{11} + \omega_{20}\hat{\sigma}_{22} \\
                            &+ g_{x} (\hat{a}^{\dagger}\hat{b} + \hat{a}\hat{b}^{\dagger}) +
                               g_{y} (\hat{b}^{\dagger}\hat{\sigma}_{01} + \hat{b}\hat{\sigma}_{10}) \\
                            &+ g_{02} (\hat{\sigma}_{20}\hat{c} + \hat{\sigma}_{02}\hat{c}^{\dagger}) +
                               g_{12} (\hat{\sigma}_{21}\hat{c} + \hat{\sigma}_{12}\hat{c}^{\dagger}),
        \end{split}
    \end{equation}
    where $\hat{\sigma}_{ij} \equiv |i \rangle\langle j|\,\,(i,j = 0,1,2)$ and $\omega_{ij}$ is the transition frequency for $|j\rangle \rightarrow |i\rangle$.
    Here, $\omega_{10}(= \Delta_{\mathrm{gs}})$ is a microwave transition frequency, whereas $\omega_{20}$ is an optical transition frequency, corresponding to the zero-phonon line (ZPL) of NV$^0$.
    The microwave cavity mode $\hat{a}$ couples to the mechanical cavity mode $\hat{b}$ via the piezoelectric effect with strength $g_x$.
    Typically, $g_x/2\pi \sim 1~\mathrm{MHz}$ in piezo-optomechanical systems \cite{Mirhosseini_Nature2020, Jiang_NatPhys2023, Meesala_NatPhys2024, Weaver_Natnano2024, Thiel_NatPhys2025, Zhao_Natnano2025}.
    The mechanical cavity mode couples to the $|0\rangle \leftrightarrow |1\rangle$ transition with strength $g_y$.
    Although the achievable mechanical coupling for NV$^0$ in an optomechanical resonator is not yet known, couplings of the order of $10~\mathrm{MHz}$ have been estimated for other color centers, such as NV$^-$ and group-IV color centers including SiV \cite{Kim_PRL2023,Joe_Nature2026}, suggesting that a conservative estimate of $g_y/2\pi \sim 1~\mathrm{MHz}$ is reasonable.
  	On the optical side, because the cavity linewidth is taken to be broader than $\omega_{10}$, the optical cavity mode $\hat{c}$, which is tuned to $\omega_{20}$, couples to both the $|0\rangle \leftrightarrow |2\rangle$ and $|1\rangle \leftrightarrow |2\rangle$ transitions.
    These transitions obey different polarization selection rules and are controllable by the external strain field.
    In this work, we focus on the strain regime where the $|0\rangle \leftrightarrow |2\rangle$ coupling exceeds the $|1\rangle \leftrightarrow |2\rangle$ coupling, taking $g_{02}/2\pi = 1~\mathrm{GHz}$ and $g_{12}/2\pi = 0.2~\mathrm{GHz}$ (see Sec.~S1.B of the Supplemental Material \cite{SM}).
    In the following, we set $\omega_a/2\pi = \omega_b/2\pi = \omega_{10}/2\pi = 10$~GHz and $\omega_{20}/2\pi = \omega_c/2\pi = 500$~THz, unless otherwise specified.

    Under the system Hamiltonian and the loss channels shown in Fig.~\ref*{FIG:schematic}(b), the Heisenberg-Langevin equation for a system operator
    $\hat{O}_{\vec{\lambda}} = (\hat{a}^{\dagger})^{p} (\hat{a})^{q} (\hat{b}^{\dagger})^{r} (\hat{b})^{s} (\hat{c}^{\dagger})^{u} (\hat{c})^{v} \hat{\sigma}_{mn}$
    is written as
    \begin{align}
        \label{eq:Heisenberg_equation}
        &\frac{d}{dt}\hat{O}_{\vec{\lambda}_1} = \sum_{\vec{\lambda}_2}
        \Big( M_{\vec{\lambda}_1,\vec{\lambda}_2}^{(1)}\hat{O}_{\vec{\lambda}_2} + 
        M_{\vec{\lambda}_1,\vec{\lambda}_2}^{(2)}\hat{a}_{\mathrm{in}}^{\dagger}(t)\hat{O}_{\vec{\lambda}_2} + \\
        &M_{\vec{\lambda}_1,\vec{\lambda}_2}^{(3)}\hat{O}_{\vec{\lambda}_2}\hat{a}_{\mathrm{in}}(t) +
        M_{\vec{\lambda}_1,\vec{\lambda}_2}^{(4)}\hat{c}_{\mathrm{in}}^{\dagger}(t)\hat{O}_{\vec{\lambda}_2} + 
        M_{\vec{\lambda}_1,\vec{\lambda}_2}^{(5)}\hat{O}_{\vec{\lambda}_2}\hat{c}_{\mathrm{in}}(t)\Big), \nonumber
    \end{align}
    where $\vec{\lambda} \equiv (p,q,r,s,u,v,m,n)$ and $p,q,r,s,u,v$ are non-negative integers.
    The coefficient tensors $M_{\vec{\lambda}_1,\vec{\lambda}_2}^{(1\sim5)}$ incorporate the unitary time evolution determined by the system Hamiltonian [Eq.~(\ref*{eq:system_Hamiltonian})] as well as the external and intrinsic loss channels characterized by $\kappa_\ell^e$ and $\kappa_\ell^i$ of mode $\ell$, and the population decay rates $\gamma_{10}$, $\gamma_{20}$ and $\gamma_{21}$.

    In the present device, we simultaneously input a classical (coherent-state) pump field from the optical waveguide and a target single photon from the microwave waveguide.
    However, instead of a single photon, we analyze the case of a coherent-state input also from the microwave waveguide, and {\it translate} the results to the case of a single-photon input \cite{KoshinoPRL2004,KoshinoPRL2007}.
    When the input fields are in coherent states, the input-field operators are rigorously replaceable with their expectation values as $\hat{a}_\mathrm{in}(t) \to \langle \hat{a}_{\mathrm{in}}(t) \rangle = E_{\mathrm{s}} e^{-i \mu_{\mathrm{s}} t}$ and $\hat{c}_{\mathrm{in}}(t) \to \langle \hat{c}_{\mathrm{in}}(t) \rangle = E_{\mathrm{p}} e^{-i \mu_{\mathrm{p}} t}$, where $E_{\mathrm{s}}$ ($E_{\mathrm{p}}$) represents the signal (pump) amplitude and $\mu_{\mathrm{s}}$ ($\mu_{\mathrm{p}}$) represents the signal (pump) frequency.
    Because $\hat{O}_{\vec{\lambda}}$ is driven by the two tones at frequencies $\mu_{\mathrm{p}}$ and $\mu_{\mathrm{s}}$, its stationary solution can be written as
    \begin{equation}
        \label{eq:stationary_solution}
        \langle \hat{O}_{\vec{\lambda}}(t) \rangle  = \sum_{\alpha,\beta} s_{\vec{\lambda}}^{(\alpha,\beta)} e^{-i(\alpha \mu_{\mathrm{p}} + \beta \mu_{\mathrm{s}})t},
    \end{equation}
    where $s_{\vec{\lambda}}^{(\alpha,\beta)}$ is a complex coefficient, and $\alpha$ and $\beta$ are integers. 
    The signal frequency $\mu_{\mathrm{s}}$, pump frequency $\mu_{\mathrm{p}}$, and target frequency $\mu_{\mathrm{p}}+\mu_{\mathrm{s}}$ correspond to $(\alpha,\beta)=(0,1)$, $(1,0)$, and $(1,1)$, respectively.
    Substituting Eq.~(\ref*{eq:stationary_solution}) into Eq.~(\ref*{eq:Heisenberg_equation}) yields simultaneous equations for $s_{\vec{\lambda}}^{(\alpha,\beta)}$.    
    
    Using $\langle \hat{O}_{\vec{\lambda}}(t) \rangle$ thus obtained and the optical input-output relation $\hat{c}_{\mathrm{out}}(t) = \hat{c}_{\mathrm{in}}(t) - i \sqrt{\kappa_{c}^{e}}\hat{c}$, we can evaluate the amplitude $\langle\hat{c}_{\mathrm{out}}(t)\rangle$ and photon flux $\langle\hat{c}_{\mathrm{out}}^{\dagger}(t)\hat{c}_{\mathrm{out}}(t)\rangle$ of the light output through the optical waveguide.
    Owing to the large frequency separation between the target and pump frequencies, the unwanted pump-frequency component can be removed using a frequency filter.
    Therefore, we define the total output photon flux $R_{\mathrm{tot}} = \langle\hat{c}_{\mathrm{out}}^{\dagger}(t)\hat{c}_{\mathrm{out}}(t)\rangle - |\langle\hat{c}_{\mathrm{out}}^{(1,0)}(t)\rangle|^2$ and the target coherent photon flux $R_{\mathrm{coh}} = |\langle\hat{c}_{\mathrm{out}}^{(1,1)}(t)\rangle|^2$.
    Here, $\langle\hat{c}_{\mathrm{out}}^{(\alpha,\beta)}(t)\rangle$ denotes the output-field-amplitude component oscillating as $\sim e^{-i(\alpha \mu_{\mathrm{p}} + \beta \mu_{\mathrm{s}})t}$.
    Note that $R_{\mathrm{tot}}$ contains incoherent emission arising from the color-center nonlinearity, while $R_{\mathrm{coh}}$ accounts only for the coherent component that preserves the coherence of the input microwave photon.

    \begin{figure}
        \centering
         \includegraphics[width=\linewidth]{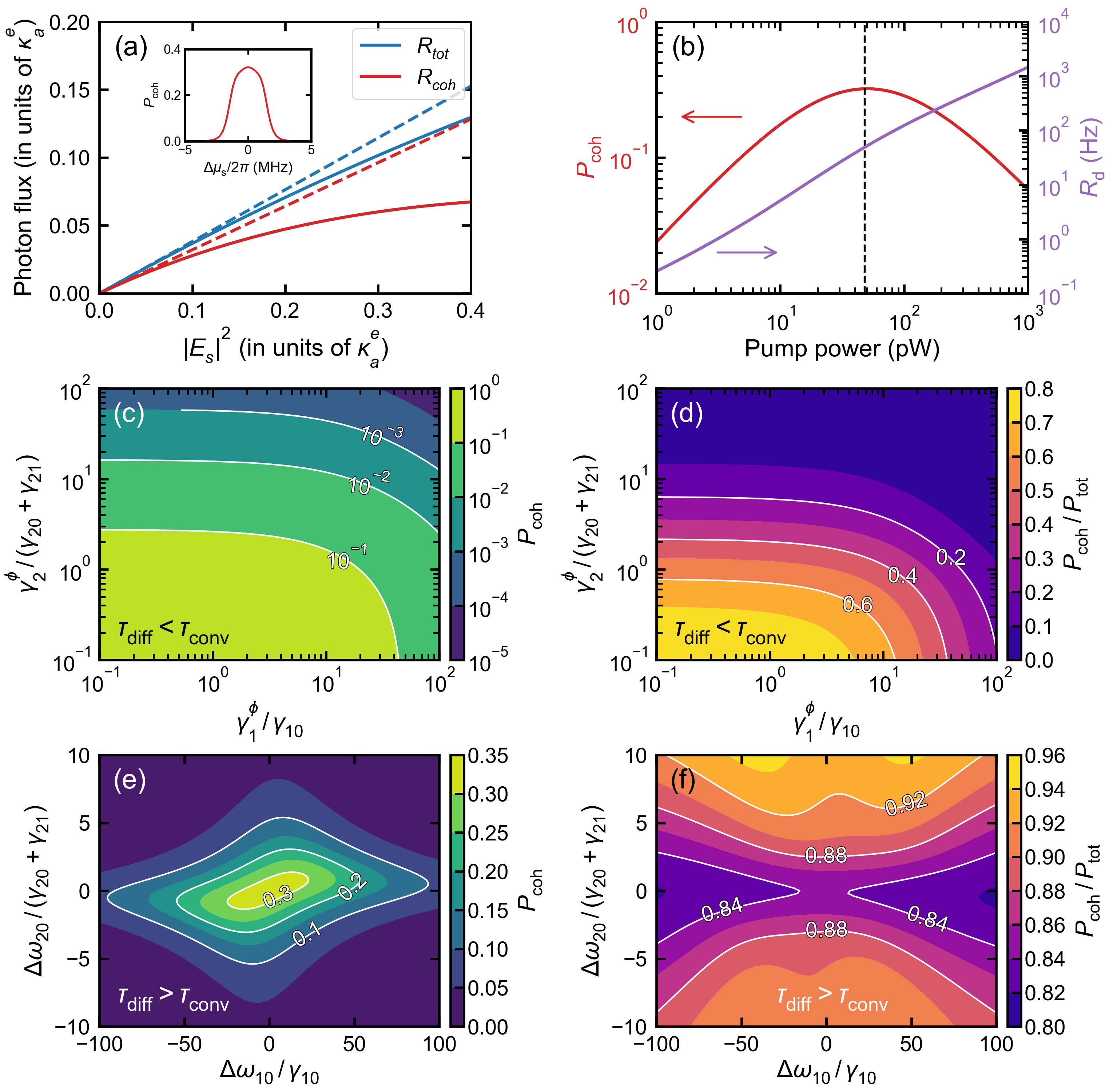}
        \caption{Single-photon conversion process. 
                 (a) Total output photon flux $R_{\mathrm{tot}}$ (blue) and its coherent component $R_{\mathrm{coh}}$ (red) as a function of signal photon flux $|E_{\mathrm{s}}|^2$.
                 The dashed lines are tangents at the origin, whose slopes give $P_{\mathrm{tot}}$ and $P_{\mathrm{coh}}$;
                 inset: $P_{\mathrm{coh}}$ versus $\Delta\mu_{\mathrm{s}}$, with the bandwidth defined as the full width at half maximum.
                 (b) Coherent conversion efficiency $P_{\mathrm{coh}}$ (red) and dark count rate (purple) as functions of pump power. The optimal pump power is 48~pW (vertical dotted line).
                 (c) $P_{\mathrm{coh}}$ and (d)  $P_{\mathrm{coh}}/P_{\mathrm{tot}}$ as functions of the pure-dephasing rates $\gamma_{1}^{\phi},\gamma_{2}^{\phi}$ for $\tau_{\mathrm{diff}} < \tau_{\mathrm{conv}}$.
                 (e) $P_{\mathrm{coh}}$ and (f) $P_{\mathrm{coh}}/P_{\mathrm{tot}}$ as functions of the detunings $\Delta\omega_{10}$ and $\Delta\omega_{20}$ for $\tau_{\mathrm{diff}} > \tau_{\mathrm{conv}}$.
                 In these panels, the signal and pump frequencies are fixed at $\omega_{10}$ and $\omega_{21}$, respectively (i.e., $\mu_{\mathrm{s}} = \omega_{10}$ and $\mu_{\mathrm{p}} = \omega_{21}$).
                } \label{FIG:1ph_conv}
    \end{figure}

	To calculate the single-photon response, we use the correspondence between coherent and number states.
    Expanding a coherent state $|\xi\rangle$ in the photon-number basis shows that the amplitude of the $|1\rangle$ component is proportional to $e^{-|\xi|^2/2}\xi$.
    Thus, in the limit of $|\xi|^2 \ll 1$, the terms proportional to $|\xi|^2$ in the expectation values for a coherent-state input correspond to the single-photon contributions.
    Accordingly, we expand $R_{\mathrm{tot}}$ and $R_{\mathrm{coh}}$ as Taylor series in the signal photon flux $|E_{\mathrm{s}}|^2$ and identify the coefficients of the linear terms as the single-photon response.
    The total detection probability $P_{\mathrm{tot}}$ and coherent conversion efficiency $P_{\mathrm{coh}}$ are therefore given by $P_{\mathrm{tot}} = \lim_{|E_{\mathrm{s}}|^2 \rightarrow 0} \frac{d R_{\mathrm{tot}}}{d |E_{\mathrm{s}}|^2}$ and $P_{\mathrm{coh}} = \lim_{|E_{\mathrm{s}}|^2 \rightarrow 0} \frac{d R_{\mathrm{coh}}}{d |E_{\mathrm{s}}|^2}$.
    
    To achieve efficient conversion, we first choose a pump power of $\hbar \mu_{\mathrm{p}} |E_{\mathrm{p}}|^2 = 48$~pW, as estimated using an analytically tractable phenomenological model \cite{SM}.
    Applying the above method, we obtain $P_{\mathrm{tot}} = 0.38$ and $P_{\mathrm{coh}} = 0.32$ from the slopes of $R_{\mathrm{tot}}$ and $R_{\mathrm{coh}}$ at the origin, respectively [Fig.~\ref*{FIG:1ph_conv}(a)].
    Their difference $P_{\mathrm{tot}} - P_{\mathrm{coh}}$ represents the detection probability of incoherent photons generated by spontaneous emission after dephasing at the NV$^0$ excited state $|2\rangle$.
    We also determine the conversion bandwidth by calculating $P_{\mathrm{coh}}$ as a function of the signal-frequency detuning $\Delta\mu_{\mathrm{s}} = \mu_{\mathrm{s}} - \omega_{10}$, obtaining $BW = 3.0$~MHz.
    The conversion bandwidth is primarily determined by the hybridized normal-mode spectrum of the microwave resonator, mechanical mode, and NV$^{0}$ ground-state transition.
    In the weak-pump limit, the upper and lower hybridized modes are separated from the central mode by approximately $\sqrt{g_x^2 + g_y^2}/2\pi = 1.4$~MHz, consistent with the numerically obtained bandwidth.
    The coherent conversion efficiency and bandwidth are both in good agreement with those predicted by the phenomenological model.
    Even in the absence of a signal, the pump weakly excites the NV$^0$ and yields a dark count rate $R_\mathrm{d} = \lim_{|E_{\mathrm{s}}|^2 \rightarrow 0} R_{\mathrm{tot}} = 48$~Hz.
    In this weak-pump regime, $R_\mathrm{d}$ increases linearly with the pump power [Fig.~\ref*{FIG:1ph_conv}(b)].

    \begin{figure}[tbp]
        \begin{center}
            \includegraphics[width=\columnwidth]{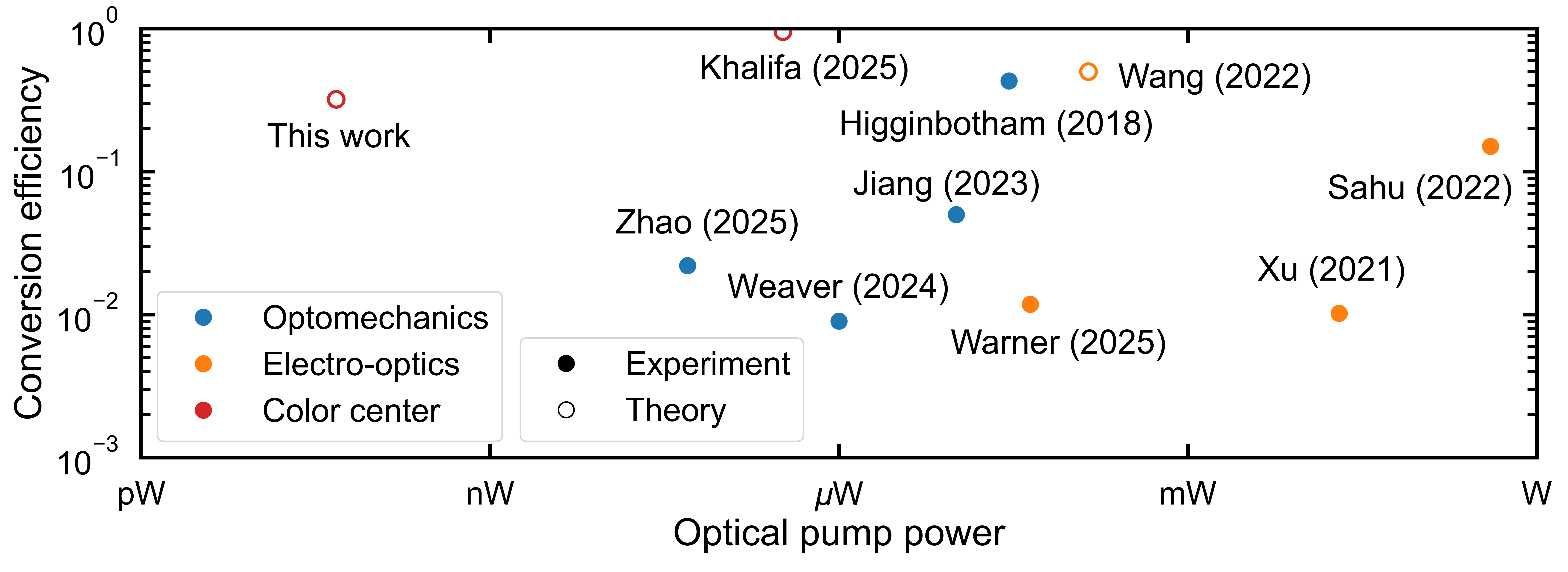}
            \caption{Conversion efficiency, including waveguide coupling efficiencies, as a function of pump power, compared with previously reported values.
                     References: Zhao (2025) \cite{Zhao_Natnano2025}; Weaver (2024) \cite{Weaver_Natnano2024}; Jiang (2023) \cite{Jiang_NatPhys2023};
                     Higginbotham (2018) \cite{Higginbotham_NatPhys2018}; Warner (2025) \cite{Warner_NatPhys2025}; Wang (2022) \cite{Wang_npj2022}; Xu (2021) \cite{Xu_NatCommun2021}; Sahu (2022) \cite{Sahu_NatCommun2022}.
                     For Ref.~\cite{Zhao_Natnano2025}, the pump power is estimated from the reported intracavity photon number;
                     all other values are taken directly from the respective references.
                    }
            \label{FIG:power_vs_efficiency}
        \end{center}
    \end{figure}

    We then compare the conversion efficiency as a function of pump power for this work and previous studies (Fig.~\ref*{FIG:power_vs_efficiency}).
    Owing to the large $g_{02}$ and the long excited-state lifetime $1 / (\gamma_{20}+\gamma_{21})$, the effective cooperativity between the optical cavity and the NV$^0$ center exceeds unity even without pump enhancement.
    Here, the cooperativity is defined as $C_{2c} = \frac{4g_{02}^{2}}{(\gamma_{20}+\gamma_{21})\kappa_{c}^{t}}$, where $\kappa_c^t = \kappa_c^e + \kappa_c^i$ is the total optical-cavity loss rate \cite{SM}.
    Consequently, the optimal pump power required to achieve a high conversion efficiency ($\sim 10\%$) is more than three orders of magnitude lower than that required in current electro-optic, optomechanical, and atomic systems.

    Next, spectral diffusion is a significant issue for color centers.
    In this paper, we distinguish two regimes based on the relative magnitudes of the spectral-diffusion timescale $\tau_{\mathrm{diff}}$ and the conversion timescale $\tau_{\mathrm{conv}}(= 300~\mathrm{ns})$ estimated from the time evolution using a simple model \cite{SM}, and analyze each case separately.
    When $\tau_{\mathrm{diff}} < \tau_{\mathrm{conv}}$, we model spectral diffusion by incorporating the pure-dephasing rates $\gamma_{1}^{\phi}$ and $\gamma_{2}^{\phi}$ of the levels $|1\rangle$ and $|2\rangle$, respectively, into $M_{\vec{\lambda}_1,\vec{\lambda}_2}^{(1)}$ in Eq.~(\ref*{eq:Heisenberg_equation}).
    The resulting coherent conversion efficiency $P_{\mathrm{coh}}$ is shown in Fig.~\ref*{FIG:1ph_conv}(c).
    To characterize the crossover from predominantly coherent to predominantly incoherent emission, we compare the pure-dephasing contribution to the decay of the optical coherence $\langle\hat{\sigma}_{02}\rangle$ with the cavity-induced contribution.
    The cavity-enhanced population decay rate through the $|2\rangle \rightarrow |0\rangle$ transition is $\Gamma_{\mathrm{cav}} = 4g_{02}^2/\kappa_c^t$, and therefore contributes $\Gamma_{\mathrm{cav}}/2$ to the decay rate of $\langle\hat{\sigma}_{02}\rangle$.
    Equating this contribution with $\gamma_{2}^{\phi}$ gives $\gamma_{2}^{\phi}/(\gamma_{20}+\gamma_{21}) = C_{2c}/2 = 1.2$.
    This rate-balance condition provides an estimate for the crossover condition $P_{\mathrm{coh}}/P_{\mathrm{tot}} = 0.5$.
    For small $\gamma_{1}^{\phi}$, the numerical calculation gives $P_{\mathrm{coh}}/P_{\mathrm{tot}} = 0.5$ at $\gamma_{2}^{\phi}/(\gamma_{20}+\gamma_{21}) = 1.3$ [Fig.~\ref*{FIG:1ph_conv}(d)], in close agreement with the above estimate.
    
    By contrast, when $\tau_{\mathrm{diff}} > \tau_{\mathrm{conv}}$, we model spectral diffusion as quasi-static frequency shifts of the levels $|1\rangle$ and $|2\rangle$ characterized by the detunings $\Delta\omega_{10}$ and $\Delta\omega_{20}$, respectively, and incorporate these detunings into $M_{\vec{\lambda}_1,\vec{\lambda}_2}^{(1)}$.
    As in the fast-diffusion regime, $P_{\mathrm{coh}}$ decreases monotonically with increasing detuning [Fig.~\ref*{FIG:1ph_conv}(e)], whereas the coherent-photon fraction $P_{\mathrm{coh}}/P_{\mathrm{tot}}$ increases as $\Delta\omega_{20}$ increases [Fig.~\ref*{FIG:1ph_conv}(f)].
    We attribute this behavior to the fact that the pump frequency $\mu_{\mathrm{p}}$ is resonant with $\omega_{21}$.
    The detuning therefore reduces the population of level $|2\rangle$, thereby preferentially suppressing incoherent emission.
    These results indicate that intentionally introducing a detuning $\Delta\omega_{20}/2\pi \sim 5(\gamma_{20}+\gamma_{21})/2\pi = 35$~MHz enables highly coherent frequency conversion with $P_{\mathrm{coh}}/P_{\mathrm{tot}} \ge 0.9$, albeit at the expense of conversion efficiency, with $P_{\mathrm{coh}} \sim 10^{-2}$.

    \begin{figure}
        \centering
        \includegraphics[width=0.85\columnwidth, trim=9mm 97mm 15mm 5mm, clip]{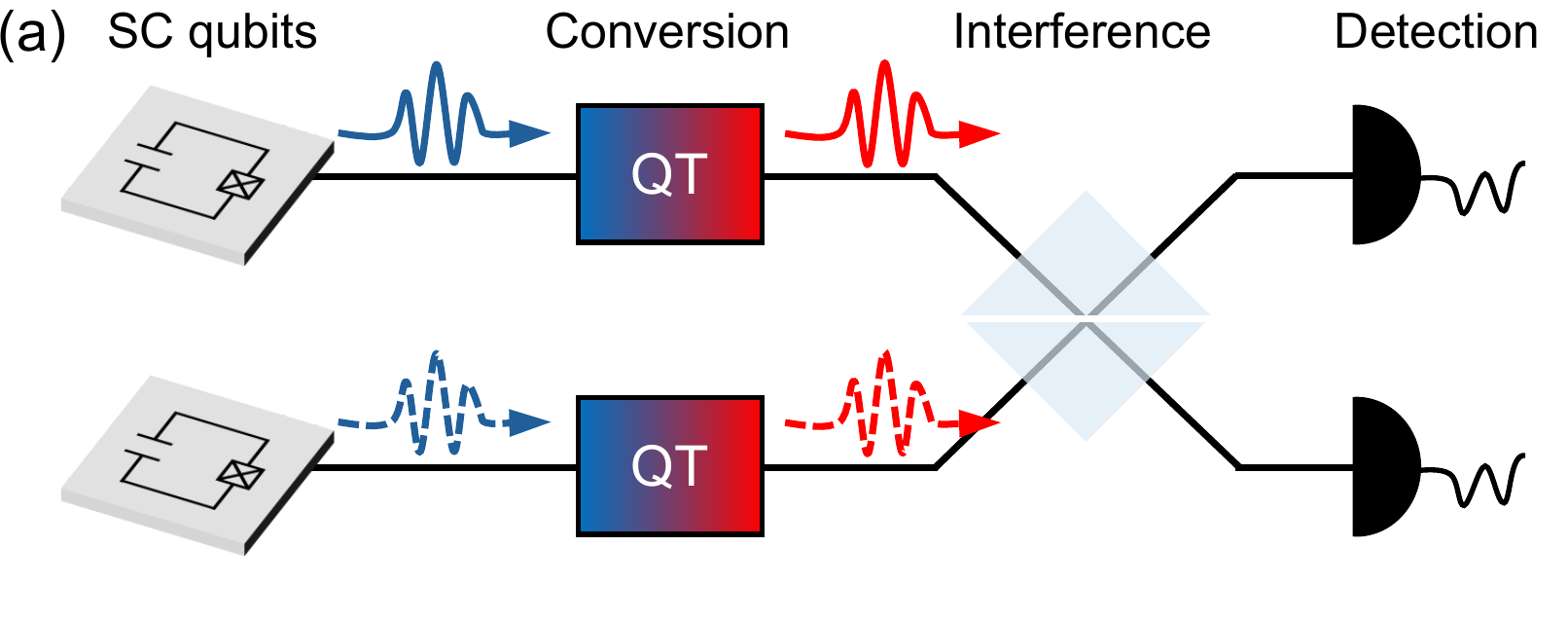}
        \vspace{3mm}
        \includegraphics[width=\columnwidth]{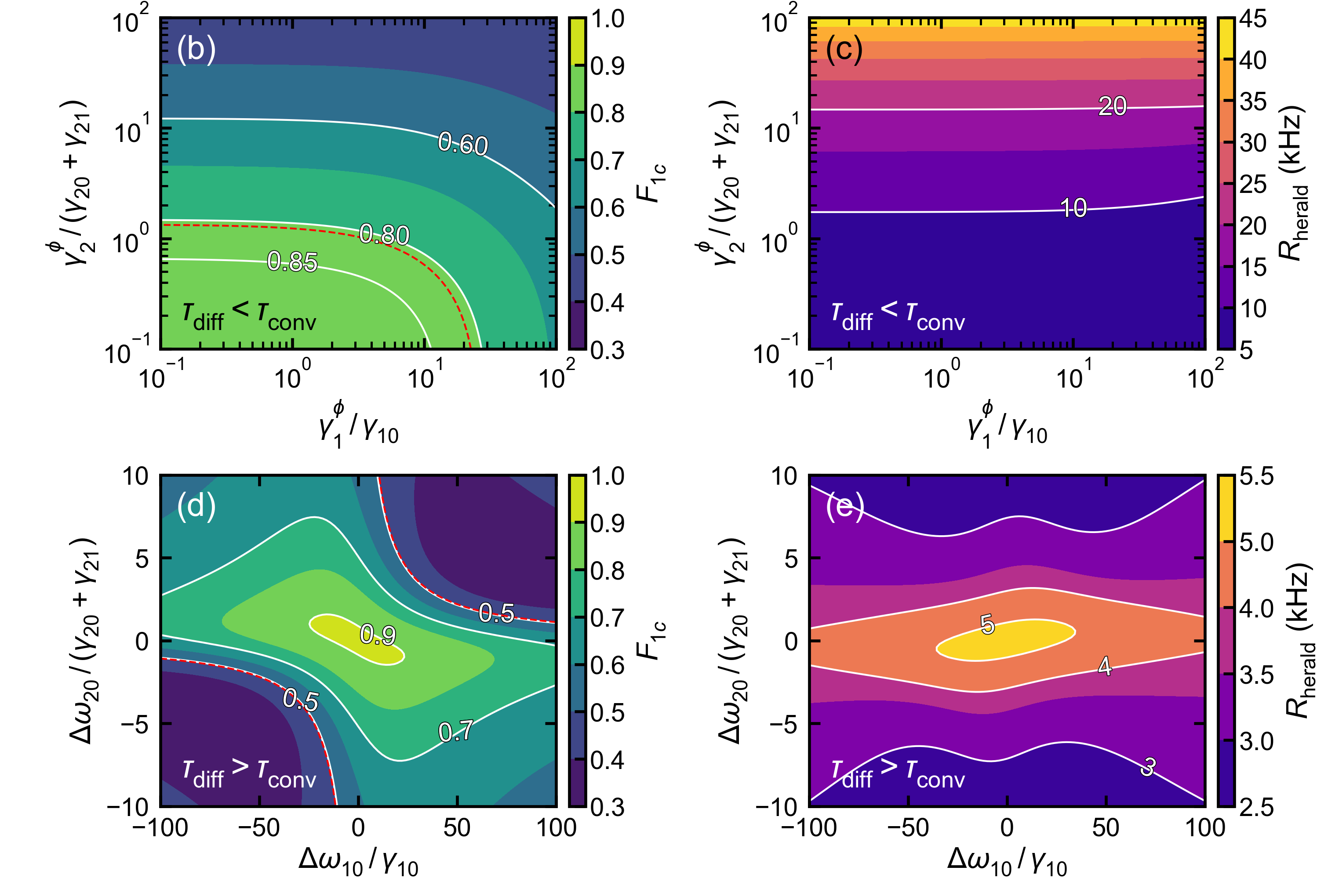}
        \caption{Remote entanglement generation. 
            (a) Microwave photons entangled with superconducting qubits (SC qubits) are converted into optical photons by quantum transducers (QTs) and injected into an interferometer, generating remote entanglement. 
            (b) Fidelity and (c) heralding rate as functions of the pure-dephasing rates $\gamma_{1}^{\phi}$ and $\gamma_{2}^{\phi}$ for $\tau_{\mathrm{diff}} < \tau_{\mathrm{conv}}$.
            The red dashed line denotes the contour $P_{\mathrm{coh}}^{(\mathrm{A})}/P_{\mathrm{tot}}^{(\mathrm{A})} = 0.5$.
            (d) Fidelity and (e) heralding rate as functions of the detunings $\Delta\omega_{10}$ and $\Delta\omega_{20}$ for $\tau_{\mathrm{diff}} > \tau_{\mathrm{conv}}$.
            The red dashed line denotes the contour $\Delta\theta = \pi/2$.
            In these panels, we set $P_{\mathrm{e}} = \sqrt{\Pi_{00}}/(\sqrt{\Pi_{00}} + \sqrt{\Pi_{11}})$, which maximizes $F_\mathrm{1c}$.
        } \label{FIG:remote_entanglement}
    \end{figure}

    Having characterized the transducer, we now consider using the converted \textit{communication} photons to generate remote entanglement between superconducting qubits located in separate dilution refrigerators.
    At each node, a photon entangled with the local superconducting qubit is frequency converted, and the converted photons from the two nodes are interfered on a beam splitter.
    In the single-click scheme considered here \cite{Cabrillo_PRA1999}, interference erases the which-node information, and a single-photon detection projects the remote qubits onto a superposition of a single excitation at either node [Fig.~\ref*{FIG:remote_entanglement}(a)].
    The entanglement fidelity is limited by imperfect photon indistinguishability caused by loss of coherence, false heralding events due to dark counts, and phase differences between the two optical paths, which alter the relative phase of the entangled state.
    High-fidelity entanglement therefore requires the qubit excitation probability $P_{\mathrm{e}}$ to be much larger than the dark-count probability $P_{\mathrm{d}}$, while remaining sufficiently small to suppress double-excitation events, i.e., $P_{\mathrm{d}} \ll P_{\mathrm{e}} \ll 1$.

    At node ${k}\,(=\mathrm{A,B})$, we model the transducer by the composite channel $\tilde{\mathcal{E}}_{k} = \mathcal{N}_{k} \circ \mathcal{E}_{k}$.
    The channel $\mathcal{E}_{k}$ converts an input photon with probability $P_{\mathrm{tot}}^{({k})}$, imparting a phase $\theta_{k}$ with coherent-conversion fraction $P_{\mathrm{coh}}^{({k})}/P_{\mathrm{tot}}^{({k})}$ conditioned on successful conversion.
    The noise channel $\mathcal{N}_{k}$ adds a noise photon with probability $P_{\mathrm{d}}^{({k})} = \tau_{\mathrm{conv}}R_{\mathrm{d}}^{({k})}$.
    Here, both the signal-pulse duration and the photon-detection window are set to $\tau_{\mathrm{conv}}$, and $\theta_{k}$ is defined as the phase-space argument of the single-photon component of $\langle\hat c_{\mathrm{out}}^{(1,1)}(t)\rangle$.
    For a link of approximately 10~m, we neglect propagation losses other than those associated with the transducers and assume phase-locked interference paths.
    The fidelity with respect to the target Bell state $|\Psi^{+}\rangle=(|10\rangle+|01\rangle)/\sqrt{2}$ is then given by
    \begin{align}
        F_{\mathrm{1c}} &= \frac{1}{N} \Big[ \Pi_{10} + \Pi_{01} +  \alpha\,\sqrt{P_{\mathrm{coh}}^{(\mathrm{A})}P_{\mathrm{coh}}^{(\mathrm{B})}}\cos{\Delta\theta}\, \Big]\,P_{\mathrm{e}}(1 - P_{\mathrm{e}}), \nonumber \\
        N               &= 2(1 - P_{\mathrm{e}})^2\,\Pi_{00} + 2P_{\mathrm{e}}^2\,\Pi_{11} \nonumber \\
                        &\quad+ 2P_{\mathrm{e}}(1 - P_{\mathrm{e}})\,\Pi_{10} + 2P_{\mathrm{e}}(1 - P_{\mathrm{e}})\,\Pi_{01}.
    \end{align}
    Here, $N$ is the total single-click probability, $\Pi_{ij}\,(i,j \in \{0,1\})$ is the single-port detection weight conditioned on the input photon-number state in the two detection paths, $\alpha = (1 - P_{\mathrm{d}}^{(A)}/2)(1 - P_{\mathrm{d}}^{(B)}/2)$, and $\Delta\theta = \theta_{\mathrm{A}} - \theta_{\mathrm{B}}$ is the phase difference between the two paths \cite{SM}.
    If entanglement generation is attempted at a repetition rate $R_{\mathrm{rep}}$, the heralding rate is approximately given by $R_{\mathrm{herald}} \sim N R_{\mathrm{rep}}$.

    Using the previously obtained results, we then evaluate the entangled-state fidelity $F_\mathrm{1c}$ and the heralding rate $R_\mathrm{herald}$ at a repetition rate of $R_{\mathrm{rep}} = 1$~MHz.
    For simplicity, spectral diffusion is assumed to occur only in the transducer at node A, while the transducer at node B is treated as ideal.
    First, when $\tau_{\mathrm{diff}} < \tau_{\mathrm{conv}}$, phase fluctuations are effectively averaged out and therefore have only a minor effect on the fidelity.
    The dominant limitation instead arises from the coherence of the emitted photons.
    Increasing $\gamma_{1}^{\phi}$ and $\gamma_{2}^{\phi}$ degrades this coherence and raises the dark-count probability $P_{\mathrm{d}}^{(\mathrm{A})}$, producing a monotonic reduction in $F_\mathrm{1c}$ [Fig.~\ref*{FIG:remote_entanglement}(b)].
    At the same time, the larger value of $P_{\mathrm{d}}^{(\mathrm{A})}$ increases the total detection probability and thereby enhances $R_\mathrm{herald}$ [Fig.~\ref*{FIG:remote_entanglement}(c)].
    When both dark-count events and phase rotations are negligible, the fidelity approximately obeys $F_{\mathrm{1c}} - 0.5 \propto \sqrt{P_{\mathrm{coh}}^{(\mathrm{A})} / P_{\mathrm{tot}}^{(\mathrm{A})}}$.
    This relation yields $F_{\mathrm{1c}} \sim 0.85$ for $P_{\mathrm{coh}}^{(\mathrm{A})}/P_{\mathrm{tot}}^{(\mathrm{A})} = 0.5$.
    Next, when $\tau_{\mathrm{diff}} > \tau_{\mathrm{conv}}$, detuning from resonance induces a substantial phase rotation and causes a pronounced reduction in the fidelity, whose behavior is governed mainly by the factor $\cos\Delta\theta$.
    As shown in Fig.~\ref*{FIG:remote_entanglement}(d), the region with $F_{\mathrm{1c}} \gtrsim 0.8$ approximately coincides with $\cos\Delta\theta \gtrsim 0.8$, while the $F_{\mathrm{1c}}=0.5$ contour nearly coincides with the contour corresponding to a phase rotation of $\pi/2$.
    The photon-detection probability, however, is largely insensitive to $\cos\Delta\theta$ and depends only weakly on detunings.
    As a result, $R_{\mathrm{herald}}$ remains nearly unchanged [Fig.~\ref*{FIG:remote_entanglement}(e)].
    Under the on-resonance condition $(\Delta\omega_{10},\Delta\omega_{20}) = (0,0)$, we obtain $F_{\mathrm{1c}}=0.91$ and $R_{\mathrm{herald}}=5.3$~kHz.

	In summary, we have proposed an extremely weak-pump microwave-to-optical quantum transducer based on a diamond optomechanical resonator containing a single NV$^0$ center.
    The proposed device yields a coherent conversion efficiency of 0.32 at a pump power of $48~\mathrm{pW}$, enabling remote entanglement generation at a rate of several kilohertz with a fidelity above $0.9$.
    While color centers offer these advantages, we also identify a challenge: spectral diffusion, for example from charge noise.
    When the spectral-diffusion timescale is shorter than the conversion timescale, incoherent emission dominates and significantly degrades the fidelity of remote entanglement generation based on photon interference.
    Thus, future research directions include mitigating such charge noise by improving fabrication conditions and exploring color centers that are robust against charge noise, as exemplified by SiV centers in diamond, while also seeking systems with small spin-orbit coupling ($<10$ GHz).	
	Despite this challenge, the proposed transducer is a promising microwave-to-optical interface and a foundation for future large-scale distributed quantum computing.

    H. Kosaka acknowledges the funding support from the Japan Science and Technology Agency (JST) Moonshot R\&D grant (JPMJMS2062) and a JST CREST grant (JPMJCR1773).
    H. Kosaka also acknowledges the Ministry of Internal Affairs and Communications (MIC) for funding the research and development to construct a global quantum cryptography network (JPMI00316), and the Japan Society for the Promotion of Science (JSPS) Grants-in-Aid for Scientific Research (20H05661, 20K20441).
    This work is also supported by the Japan Science and Technology Agency (JST) as part of Adopting Sustainable Partnerships for Innovative Research Ecosystems (ASPIRE) (JPMJAP24C1).
    
\bibliography{reference_PRL_arxiv}

% ============================================================
% Supplemental Material
% ============================================================
\clearpage
\onecolumngrid
\setcounter{secnumdepth}{3}

% Reset counters and add S-prefixes.
\setcounter{section}{0}
\setcounter{subsection}{0}
\setcounter{subsubsection}{0}
\setcounter{equation}{0}
\setcounter{figure}{0}
\setcounter{table}{0}

\renewcommand{\thesection}{S\arabic{section}}
\renewcommand{\theequation}{S\arabic{equation}}
\renewcommand{\thefigure}{S\arabic{figure}}
\renewcommand{\thetable}{S\arabic{table}}

\begin{center}
{\large\bfseries Supplemental Material for\\[0.3em]
Ultra-Weak-Pump Microwave-to-Optical Quantum Transduction via a Single Color Center}
\end{center}

\tableofcontents
\clearpage

\section{Parameters} \label{sec:parameters}

    In this section, we describe the parameter settings used in this study, as summarized in Table~\ref{tab:parameters}.
    Although color centers coupled simultaneously to both phononic and photonic crystal cavities have been investigated in several previous studies, their coupling strength has not, to our knowledge, been reported for the NV$^0$ center considered here.
    We therefore discuss the cavity-NV$^0$ coupling strength in detail below.

    \subsection{Parameter settings}

        \begin{table}[b]
            \centering
            \begin{minipage}{0.55\textwidth}
                \caption{System parameter settings in the main text.
                    Parameters marked with $^\dagger$ are discussed in this section.
                    Parameters marked with $^{\dagger\dagger}$ are chosen to optimize the conversion efficiency within the phenomenological model, as discussed in Sec.~\ref{subsec:numerical_results}.} \label{tab:parameters}
                \centering
                \begin{ruledtabular}
                    \begin{tabular}{lrc}
                        \textrm{Parameter} & \textrm{Value} & \textrm{Reference}\\
                        \hline
                        
                        \multicolumn{2}{l}{\textit{Frequencies}}\\
                        \textrm{Ground-state splitting, $\omega_{10}/2\pi$}                               & $10~\mathrm{GHz}$    & \cite{Baier2020,Kurokawa2024,Kurokawa2025} \\
                        \textrm{Optical transition frequency, $\omega_{20}/2\pi$}                         & $500~\mathrm{THz}$   & \cite{Baier2020,Kurokawa2024,Kurokawa2025}\\
                        \hline
                        
                        \multicolumn{2}{l}{\textit{Coupling rates}}\\
                        \textrm{Microwave-mechanical coupling rate, $g_x/2\pi$}                           & $1~\mathrm{MHz}$     & \cite{Mirhosseini_Nature2020, Jiang_NatPhys2023, Meesala_NatPhys2024, Weaver_Natnano2024, Thiel_NatPhys2025, Zhao_Natnano2025, Kim_PRL2023}\\
                        \textrm{Mechanical-NV$^{0}$ coupling rate, $g_y/2\pi$}                            & $1~\mathrm{MHz}$     & \cite{Kim_PRL2023,Joe_Nature2026} \\
                        \textrm{Optical $|0\rangle\leftrightarrow|2\rangle$ coupling rate, $g_{02}/2\pi$} & $1~\mathrm{GHz}$     & $^\dagger$ \\
                        \textrm{Optical $|1\rangle\leftrightarrow|2\rangle$ coupling rate, $g_{12}/2\pi$} & $0.2~\mathrm{GHz}$   & $^\dagger$ \\
                        \hline
                        
                        \multicolumn{2}{l}{\textit{Loss and decay rates}}\\
                        \textrm{Microwave internal loss rate, $\kappa_a^{i}/2\pi$}                        & $0.5~\mathrm{MHz}$   & \cite{Mirhosseini_Nature2020, Jiang_NatPhys2023, Meesala_NatPhys2024, Weaver_Natnano2024, Thiel_NatPhys2025, Zhao_Natnano2025} \\
                        \textrm{Microwave external coupling rate, $\kappa_a^{e}/2\pi$}                    & $2.15~\mathrm{MHz}$   & $^{\dagger\dagger}$ \\
                        \textrm{Mechanical internal loss rate, $\kappa_b^{i}/2\pi$}                       & $0.5~\mathrm{MHz}$   & \cite{Joe_Nature2026,Oh_Optica2026} \\
                        \textrm{Optical internal loss rate, $\kappa_c^{i}/2\pi$}                          & $50~\mathrm{GHz}$    & \cite{Joe_Nature2026,Oh_Optica2026} \\
                        \textrm{Optical external coupling rate, $\kappa_c^{e}/2\pi$}                      & $175~\mathrm{GHz}$   & $^{\dagger\dagger}$ \\
                        \textrm{$|1\rangle\to|0\rangle$ decay rate, $\gamma_{10}/2\pi$}                   & $33~\mathrm{kHz}$    & \cite{Kurokawa2025} \\
                        \textrm{$|2\rangle\to|0\rangle$ decay rate, $\gamma_{20}/2\pi$}                   & $3.5~\mathrm{MHz}$   & \cite{Baier2020} \\
                        \textrm{$|2\rangle\to|1\rangle$ decay rate, $\gamma_{21}/2\pi$}                   & $3.5~\mathrm{MHz}$   & \cite{Baier2020} \\
                    \end{tabular}
                \end{ruledtabular}
            \end{minipage}
        \end{table}

        \subsubsection{Frequencies and decay rates}
        Because our quantum transducer is based on an architecture similar to that of conventional transducers employing optomechanical crystal cavities, most resonator parameters can be taken from previous studies.
        For the microwave resonator, we use parameters reported in studies on silicon optomechanical systems \cite{Mirhosseini_Nature2020, Jiang_NatPhys2023, Meesala_NatPhys2024, Weaver_Natnano2024, Thiel_NatPhys2025, Zhao_Natnano2025}.
        For the diamond optomechanical resonator, the available literature on similar structures is more limited than that for silicon systems; we therefore refer to Refs.~\cite{Joe_Nature2026,Oh_Optica2026}.
        The material parameters of the NV$^0$ center are taken from Refs.~\cite{Baier2020,Kurokawa2024,Kurokawa2025}.

        For numerical simplicity, we set $\omega_{10}/2\pi = 10$~GHz and use $\omega_{20}/2\pi = 500$~THz as an approximate value for the NV$^0$ zero-phonon line (ZPL) frequency throughout the transduction calculations.
        We note that the strain regime used to obtain the assumed polarization asymmetry would lead to a larger ground-state splitting, of order 20~GHz.
        This difference does not qualitatively affect the main results, such as the conversion efficiency, optimal pump power, and entanglement fidelity.

        \subsubsection{Coupling strengths}
        The coupling strengths between the relevant modes are chosen as follows.
        For the microwave-mechanical coupling strength, we use experimental values reported for silicon optomechanical systems \cite{Mirhosseini_Nature2020, Jiang_NatPhys2023, Meesala_NatPhys2024, Weaver_Natnano2024, Thiel_NatPhys2025, Zhao_Natnano2025} and calculated values for diamond optomechanical resonators \cite{Kim_PRL2023}.
        For the coupling strength between the mechanical mode and the color center, we refer to Refs.~\cite{Kim_PRL2023,Joe_Nature2026}.
        The coupling strength between the color center and the optical cavity is estimated based on the calculation described in the next subsection.

        We note that previous studies of mechanical-mode-color-center coupling \cite{Kim_PRL2023,Joe_Nature2026} considered the excited state of the NV$^-$ center and the SiV$^-$ center, respectively.
        However, the ground state of the NV$^0$ center possesses orbital degrees of freedom similar to those in these systems.
        We therefore assume a comparable strain susceptibility of order 1~PHz per unit strain \cite{Peng_npj2025}.
        Based on this assumption, we adopt a conservative estimate for the mechanical-mode-NV$^0$ coupling strength.

        \subsubsection{Frequency tunability}
        To achieve efficient conversion, we operate the system in the rotating frame at the pump frequency $\mu_{\mathrm{p}}$, where all modes are resonant; that is, $\omega_a = \omega_b = \omega_{10} = \omega_{20} - \mu_{\mathrm{p}} = \omega_c - \mu_{\mathrm{p}}$.
        To this end, the resonance condition must be satisfied by tuning the frequencies of the individual resonators, which are subject to fabrication-related variations.

        For the microwave resonator coupled to the optomechanical cavity \cite{Arriola_PRX2018}, the frequency can be tuned by applying a magnetic field to a SQUID \cite{Sandberg_AppPhyisLett2008} or a high-kinetic-inductance ($L_k$) ladder resonator \cite{Xu_AppPhyisLett2019}.
        For the mechanical resonator, its frequency is assumed to be fixed.
        For the NV$^0$ center, the frequency can be controlled by locally applying an electric field \cite{Kurokawa2024}.
        For the optical resonator, the frequency can be tuned by attaching nitrogen gas to the photonic crystal, thereby changing its refractive index \cite{Nguyen_PRB2019}.
    
    \subsection{Coupling strength between the \texorpdfstring{NV$^0$}{NV0} center and the optical cavity}
        \subsubsection{Energy levels of \texorpdfstring{NV$^0$}{NV0}}
        As mentioned in the main text, an NV$^0$ center has four ground states formed by the orbital states ($|\pm\rangle_{\mathrm{o}} = 1/\sqrt{2} (|e_x\rangle_{\mathrm{o}} \mp i |e_y\rangle_{\mathrm{o}})$) and the spin states ($|\uparrow\rangle_{\mathrm{s}},|\downarrow\rangle_{\mathrm{s}}$) \cite{Baier2020,Kurokawa2024,Kurokawa2025}.
        Here, $|e_x\rangle_{\mathrm{o}}$ and $|e_y\rangle_{\mathrm{o}}$ are the orbital eigenstates of NV$^0$.
        We define the orbital operators $\hat{L}_{z} = |+\rangle_{\mathrm{o}}\langle+| - |-\rangle_{\mathrm{o}}\langle-|$ and $\hat{L}_{\pm} = |\pm\rangle_{\mathrm{o}}\langle \mp|$ and the half-spin operator $\hat{S}_{z} = (|\uparrow\rangle_{\mathrm{s}}\langle \uparrow| - |\downarrow\rangle_{\mathrm{s}}\langle \downarrow|)/2$.
        The ground-state Hamiltonian of NV$^0$ under a zero magnetic field is then given by
        \begin{equation}
            \label{eq:ground_state_Hamiltonian}
            \hat{H}_{\mathrm{NV^0}}/\hbar = 2\lambda_{\mathrm{so}}\hat{L}_{z}\hat{S}_{z} + \varepsilon_{\perp}(\hat{L}_{+}+\hat{L}_{-}),
        \end{equation}
        where $\lambda_{\mathrm{so}}$ is the spin-orbit interaction parameter and $\varepsilon_{\perp}$ is the strain parameter perpendicular to the NV axis.

        The ground-state Hamiltonian in Eq.~(\ref{eq:ground_state_Hamiltonian}) is diagonalized by the following basis states:
        \begin{align}
            |\pm'\rangle_{\mathrm{o}}|\uparrow\rangle_{\mathrm{s}} &= \frac{ (\lambda_{\pm} |+\rangle_{\mathrm{o}} + \varepsilon_{\perp} |-\rangle_{\mathrm{o}}) |\uparrow\rangle_{\mathrm{s}} }{ \sqrt{ \lambda_{\pm}^2 + \varepsilon_{\perp}^2 } }, \label{NV_eigenstates_with_strain_1} \\
            |\pm'\rangle_{\mathrm{o}}|\downarrow\rangle_{\mathrm{s}} &= \frac{ (-\lambda_{\pm} |+\rangle_{\mathrm{o}} + \varepsilon_{\perp} |-\rangle_{\mathrm{o}}) |\downarrow\rangle_{\mathrm{s}} }{ \sqrt{ \lambda_{\pm}^2 + \varepsilon_{\perp}^2 } }, \label{NV_eigenstates_with_strain_2}
        \end{align}
        where $\lambda_{\pm} = \lambda_{\mathrm{so}} \pm \sqrt{\lambda_{\mathrm{so}}^2 + \varepsilon_{\perp}^2}$.
        The ground states are therefore split into two spin-degenerate branches separated by $\Delta_{\mathrm{gs}} = 2\sqrt{\lambda_{\mathrm{so}}^2 + \varepsilon_{\perp}^2}$.
        The lower branch consists of $|-'\rangle_{\mathrm{o}}|\uparrow\rangle_{\mathrm{s}}$ and $|+'\rangle_{\mathrm{o}}|\downarrow\rangle_{\mathrm{s}}$, whereas the upper branch consists of $|+'\rangle_{\mathrm{o}}|\uparrow\rangle_{\mathrm{s}}$ and $|-'\rangle_{\mathrm{o}}|\downarrow\rangle_{\mathrm{s}}$.
        On the other hand, the orbital excited state, $|0\rangle_{\mathrm{o}}|\uparrow\rangle_{\mathrm{s}}$ and $|0\rangle_{\mathrm{o}}|\downarrow\rangle_{\mathrm{s}}$, has no spin-orbit splitting, and the spin states are degenerate under a zero magnetic field \cite{Baier2020, Barson_Nanophotonics2018}. 
        In the main text, we use $|0\rangle$, $|1\rangle$, and $|2\rangle$ to denote, respectively, the lower ground-state manifold, the upper ground-state manifold, and the excited-state manifold assuming a zero magnetic field. 
        
        \subsubsection{Polarization selection rules for excited-state transitions}
        In the main text, we assume an imbalance in the NV-optical coupling such that $g_{02}>g_{12}$.
        This asymmetric coupling between the transition dipoles and the cavity electric field can arise under suitable strain conditions, where the two dipole transitions, $\ket{0}\rightarrow\ket{2}$ and $\ket{1}\rightarrow\ket{2}$, obey different polarization selection rules, as shown below.
        
        We discuss the polarization selection rules by focusing on the orbital transitions between the ground states $|\pm\rangle_{\mathrm{o}}$ and the excited state $|0\rangle_{\mathrm{o}}$.
        In this work, we take the horizontal ($\ket{H}_\mathrm{pol}$) and vertical ($\ket{V}_\mathrm{pol}$) polarization states to be associated with the transitions $|e_x\rangle_{\mathrm{o}} \leftrightarrow |0\rangle_{\mathrm{o}}$ and $|e_y\rangle_{\mathrm{o}} \leftrightarrow |0\rangle_{\mathrm{o}}$, respectively.
        With this convention, the right- and left-circular polarization states are defined as
        \begin{align}
            |R\rangle_{\mathrm{pol}} = \frac{1}{\sqrt{2}} (|H\rangle_{\mathrm{pol}} - i |V\rangle_{\mathrm{pol}}) \quad \mathrm{for} \quad |-\rangle_{\mathrm{o}} \leftrightarrow |0\rangle_{\mathrm{o}}, \\
            |L\rangle_{\mathrm{pol}} = \frac{1}{\sqrt{2}} (|H\rangle_{\mathrm{pol}} + i |V\rangle_{\mathrm{pol}}) \quad \mathrm{for} \quad |+\rangle_{\mathrm{o}} \leftrightarrow |0\rangle_{\mathrm{o}}.
        \end{align}

        Using these correspondences, the energy eigenstates, given in Eqs.~(\ref{NV_eigenstates_with_strain_1}) and (\ref{NV_eigenstates_with_strain_2}), can be mapped onto the corresponding polarization components as
        \begin{align}\label{eq:strain-pol}
            |\pm'\rangle_{\mathrm{o}}|\uparrow\rangle_{\mathrm{s}} &\mapsto \frac{ \Big\{(\lambda_{\pm} + \varepsilon_{\perp}) |H\rangle_{\mathrm{pol}} + i (\lambda_{\pm} - \varepsilon_{\perp}) |V\rangle_{\mathrm{pol}} \Big\} |\uparrow\rangle_{\mathrm{s}} }{ \sqrt{ 2(\lambda_{\pm}^2 + \varepsilon_{\perp}^2) } }, \\\label{eq:strain-pol2}
            |\pm'\rangle_{\mathrm{o}}|\downarrow\rangle_{\mathrm{s}} &\mapsto \frac{ \Big\{-(\lambda_{\pm} - \varepsilon_{\perp}) |H\rangle_{\mathrm{pol}} - i (\lambda_{\pm} + \varepsilon_{\perp}) |V\rangle_{\mathrm{pol}} \Big\} |\downarrow\rangle_{\mathrm{s}} }{ \sqrt{ 2(\lambda_{\pm}^2 + \varepsilon_{\perp}^2) } }.
        \end{align}

\begin{figure}
        	\centering
        	\includegraphics[width=0.8\columnwidth]{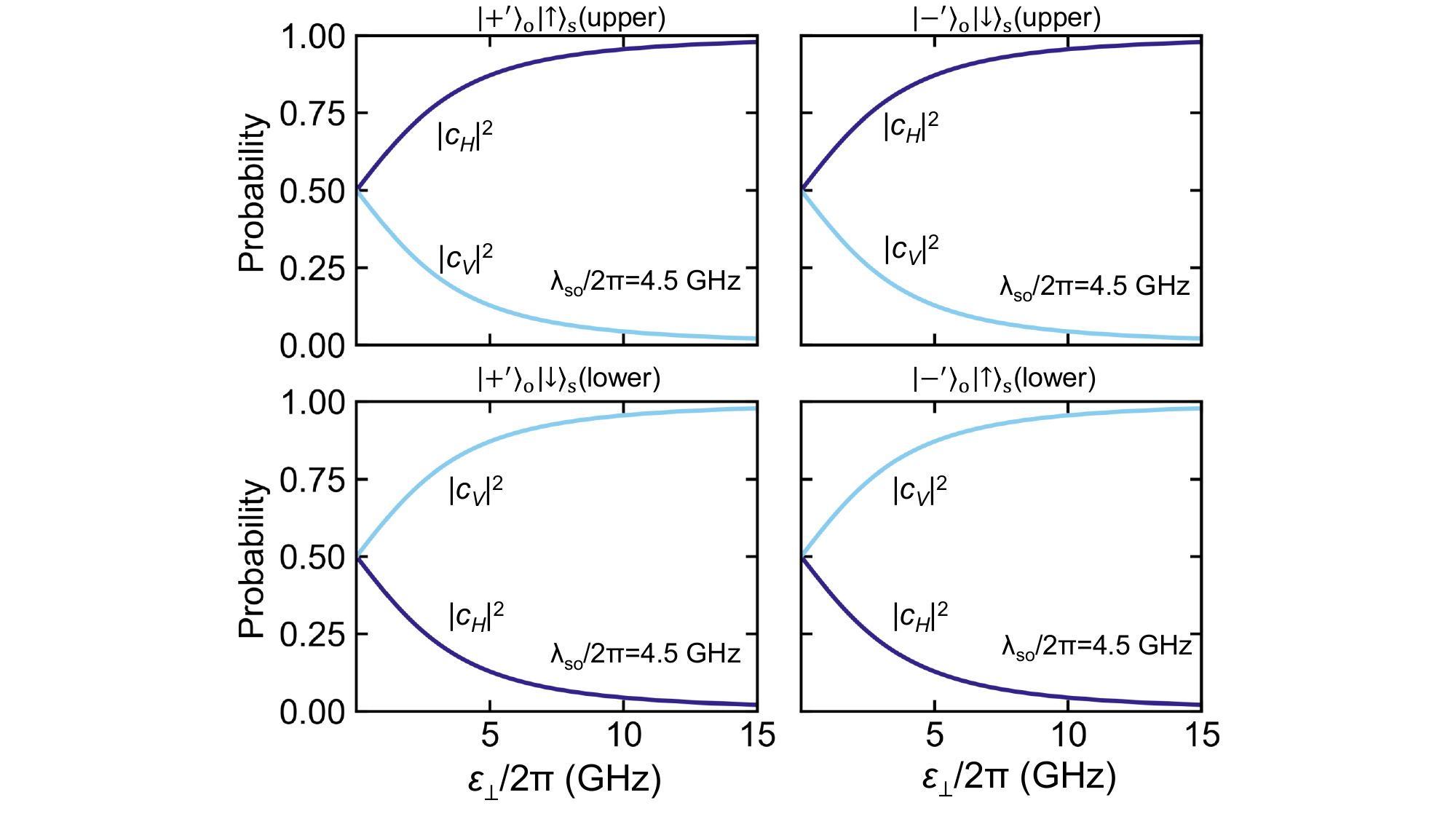}
        	\caption{Relationship between strain and polarization probability obtained from Eqs. (\ref{eq:strain-pol}) and (\ref{eq:strain-pol2}). $c_\mathrm{H}$ and $c_\mathrm{V}$ denote the coefficients in  Eqs. (\ref{eq:strain-pol}) and (\ref{eq:strain-pol2}). 
        	} \label{fig:strain-pol_probability}
        \end{figure}

\begin{figure}
        	\centering
        	\includegraphics[width=0.7\columnwidth]{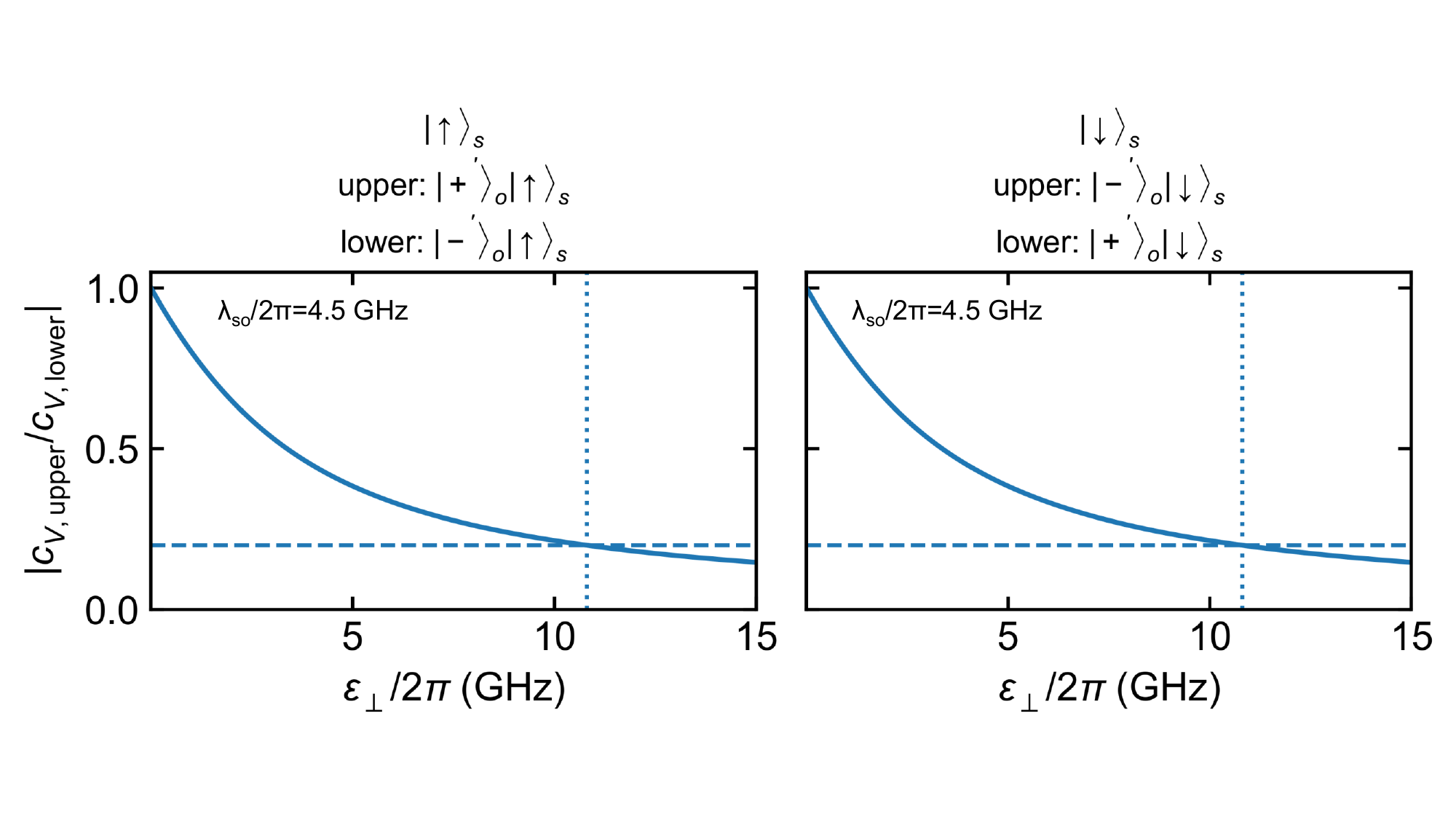}
        	\caption{Polarization ratio between  the lower and upper branches with the same spin obtained from Eqs. (\ref{eq:strain-pol}) and (\ref{eq:strain-pol2}). $c_\mathrm{V,lower}$ and $c_\mathrm{V,upper}$ denote the coefficients of $V_\mathrm{pol}$ in  Eqs. (\ref{eq:strain-pol}) and (\ref{eq:strain-pol2}). 
        	} \label{fig:strain-pol}
        \end{figure}

        Thus, the optical coupling amplitudes of the $|0\rangle \leftrightarrow |2\rangle$ and $|1\rangle \leftrightarrow |2\rangle$ transitions to a fixed cavity polarization can be tuned by the strain-dependent polarization components of the ground-state eigenstates.
        As shown in Fig.~\ref{fig:strain-pol_probability}, with increasing strain, the transition dipole of the lower (upper) branch approaches $\ket{V}_\mathrm{pol}$ ($\ket{H}_\mathrm{pol}$).
        For $\varepsilon_\perp/2\pi \sim 10$~GHz, which is approximately twice the spin-orbit coupling parameter $\lambda_\mathrm{so}/2\pi=4.5$~GHz, the coupling ratio to a $V$-polarized optical resonator becomes $g_{12}/g_{02}\propto |c_\mathrm{V,upper}/c_\mathrm{V,lower}|\sim 0.2$ [Fig.~\ref{fig:strain-pol}], where $c_\mathrm{V,upper}$ and $c_\mathrm{V,lower}$ are the coefficients of $\ket{V}_\mathrm{pol}$ in the upper and lower branches in Eqs.~\eqref{eq:strain-pol} and \eqref{eq:strain-pol2}.
        We assume this strain regime for the main-text parameters and set $g_{12} = 0.2 g_{02}$.
        
        \subsubsection{Vacuum coupling strength}
        The vacuum coupling strength between the optical cavity and an NV center is given in terms of the transition dipole moment $\vec{d}$ and the zero-point electric-field amplitude $\vec{E}^{\mathrm{zpf}}$ as
        \begin{equation}
            \label{eq:vacuum_coupling_strength}
            g_0 = \frac{\vec{d} \cdot \vec{E}^{\mathrm{zpf}}}{\hbar}.
        \end{equation}
        When the $z$ axis is chosen to be parallel to the NV axis, the transition dipole moment has only components perpendicular to the NV axis.
        Thus, we take the $x$ and $y$ components to correspond to the transitions driven by horizontal and vertical polarizations, respectively.
        In addition, we approximate the cavity field as a transverse-electric mode whose electric field is dominated by the $E_x$ component transverse to the waveguide direction.
        Eq.~(\ref{eq:vacuum_coupling_strength}) therefore reduces to $g_0 \simeq d_x E_x^{\mathrm{zpf}} / \hbar$.
        If the transition dipole moment is aligned with the $x$ direction, this expression can be written as $g_0 \simeq |\vec{d}| |\vec{E}^{\mathrm{zpf}}|/\hbar$.

        To calculate $g_0$, we estimate the transition dipole moment $|\vec{d}|$ from the Einstein coefficient. 
        In general, the radiative lifetime $\tau_0$ associated with a transition dipole moment $\vec{d}$ is given by
        \begin{equation}
            \label{eq:spontaneous_emission_rate}
            \frac{1}{\tau_0} = \frac{n\omega_0^3}{3 \pi \varepsilon_0 \hbar c^3} |\vec{d}|^2,
        \end{equation}
        where $n$ is the refractive index of the material, $\omega_0/2\pi$ is the transition frequency, $\varepsilon_0$ is the vacuum permittivity and $c$ is the speed of light in vacuum.

        \begin{table}
            \centering
            \begin{minipage}{0.55\textwidth}
                \caption{Total transition dipole moment $|\vec{d}|$ and ZPL-only moment $d_{\mathrm{ZPL}}$ estimated from $(\lambda_0, \tau_0)$ using the refractive index $n = 2.4$ for homogeneous diamond.
                The parameter sets $(\lambda_0, \tau_0)$ are taken from Refs.~\cite{Baier2020} and~\cite{Doherty_2013} for NV$^0$ and NV$^-$, respectively.
                The Debye--Waller (DW) factors are obtained from the Huang--Rhys factor reported in Ref.~\cite{Gergo_JAP2024}.
                } \label{tab:transition_moments}
                \begin{ruledtabular}
                    \begin{tabular}{cccccc}
                        \textrm{Center} & \textrm{$\lambda_0$ (nm)} & \textrm{$\tau_0$ (ns)} & \textrm{$|\vec{d}|$ (D)} & \textrm{DW} & \textrm{$d_{\mathrm{ZPL}}$ (D)} \\
                        \hline
                        NV$^0$ & 575 & 22 & 3.39 & 0.10 & 1.071 \\
                        NV$^-$ & 637 & 13 & 5.14 & 0.03 & 0.890 \\
                    \end{tabular}
                \end{ruledtabular}
            \end{minipage}
        \end{table}

        Using Eq.~(\ref{eq:spontaneous_emission_rate}), we estimate the magnitudes of the optical transition dipole moments of the NV$^0$ and NV$^-$ centers from the zero-phonon-line (ZPL) wavelengths $\lambda_0$ and the excited-state lifetimes $\tau_0$ in bulk diamond.
        Since we focus on the ZPL transitions, we introduce the ZPL-only transition dipole moment as $d_{\mathrm{ZPL}} = \sqrt{\mathrm{DW}}\,|\vec{d}|$ where DW denotes the Debye--Waller factor.
        Table~\ref{tab:transition_moments} summarizes $\lambda_0$, $\tau_0$, the DW factor and the corresponding transition moments for NV$^0$ and NV$^-$.
        Moreover, the quantum efficiency $\eta_{\mathrm{QE}}$ can also be included as an additional correction factor, $d_{\mathrm{ZPL}}(\eta_{\mathrm{QE}}) = \sqrt{\eta_{\mathrm{QE}}}\,d_{\mathrm{ZPL}}$.

        On the other hand, the zero-point electric-field amplitude of the optical cavity mode is given by
        \begin{equation}
            |\vec{E}^{\mathrm{zpf}}| = \sqrt{ \frac{\hbar \omega_c}{2 V_{\mathrm{mode}} \varepsilon_0 \varepsilon_r} },
        \end{equation}
        where $\omega_c/2\pi$ is the optical cavity resonance frequency, $V_{\mathrm{mode}}$ is the mode volume and $\varepsilon_r$ is the relative permittivity of the cavity material.
        Using $\omega_c = 2\pi c / \lambda_0$ and $V_{\mathrm{mode}} \sim (\lambda_0/n)^{3}$, with $\varepsilon_r = 5.6$ for diamond, we obtain
        \begin{equation}
            |\vec{E}^{\mathrm{zpf}}| \simeq 
            \begin{cases}
                5.0 \times 10^5~\mathrm{V/m} & \mathrm{for \quad NV^0}, \\
                4.1 \times 10^5~\mathrm{V/m} & \mathrm{for \quad NV^-}.
            \end{cases}
        \end{equation}

\begin{figure}
            \centering
            \includegraphics[width=0.45\columnwidth]{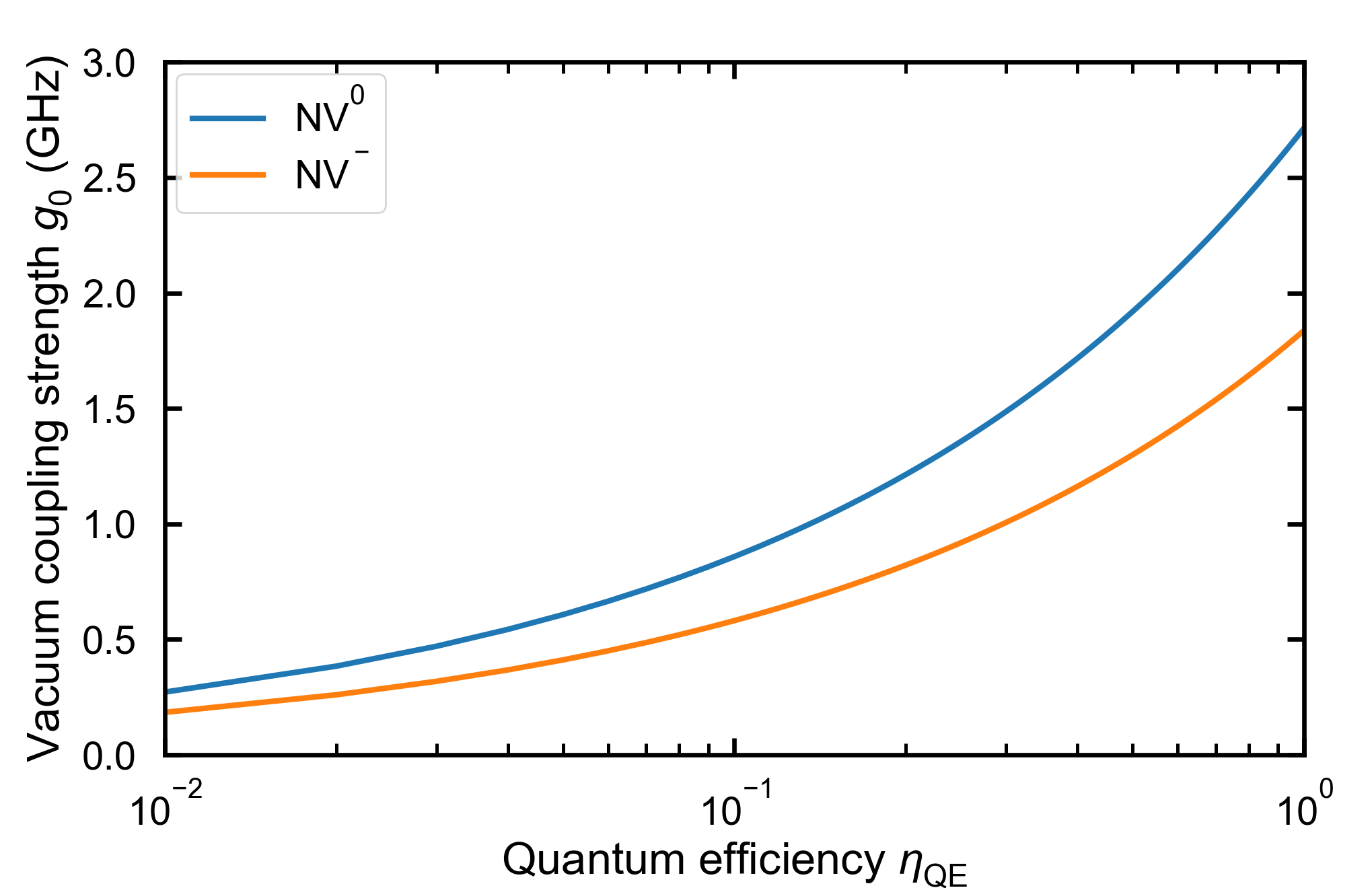}
            \caption{Vacuum coupling strengths between the optical cavity and NV$^0$ and NV$^-$ centers as a function of quantum efficiency $\eta_{\mathrm{QE}}$.
            } \label{fig:vacuum_coupling_strength}
        \end{figure}

        Finally, we define the vacuum coupling strength associated with the ZPL transition as $g_{02} = d_{\mathrm{ZPL}}(\eta_{\mathrm{QE}}) |\vec{E}^{\mathrm{zpf}}| / \hbar$ and evaluate it for various quantum efficiencies, as shown in Fig.~\ref{fig:vacuum_coupling_strength}.
        Although the quantum efficiency of NV$^0$ has been estimated to approach unity \cite{Fraczek_2017}, which would give $g_{02}/2\pi \sim 3$~GHz, nanofabrication-induced damage can reduce $\eta_\mathrm{QE}$.
        In addition, misalignment between the emission dipole and the cavity electric field can decrease the coupling strength.
        Therefore, to account for these realistic constraints, we set $g_{02}/2\pi = 1$~GHz in this work.

\section{Theoretical models of the quantum transducer}

    In this study, we analyzed three types of theoretical models of our quantum transducer: the full, reduced and phenomenological models.
    Because each model has different advantages and limitations in terms of computational cost and approximation accuracy, we employ these models selectively in the present study according to the purpose of the analysis.

    \subsection{Full model}
        The minimal theoretical model of the proposed quantum transducer can be described as a system consisting of four quantum elements coupled in series, as shown in Fig.~\ref{FIG:full_model}.
        In this work, we refer to this model as the full model to distinguish it from the reduced and phenomenological models.

        \subsubsection{Hamiltonian}

            \begin{figure}[htbp]
                \begin{center}
                \includegraphics[width=0.65\columnwidth]{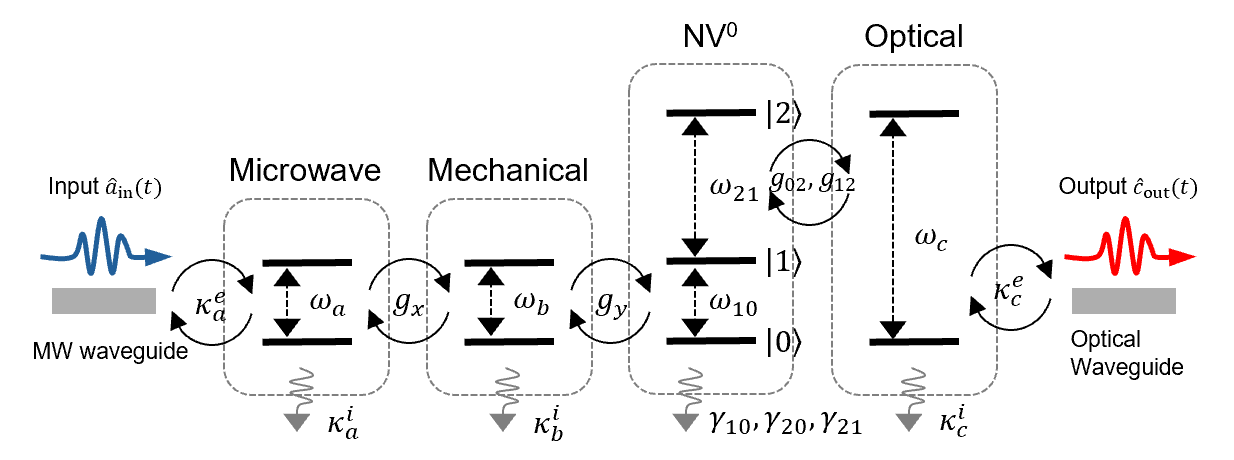}
                \caption{
                    Mode-coupling diagram.
                    We operate under the triply resonant condition in which the microwave mode, the mechanical mode, and the $|0\rangle \leftrightarrow |1\rangle$ transition satisfy $\omega_a = \omega_b = \omega_{10}$ and the optical mode is tuned to the $|0\rangle \leftrightarrow |2\rangle$ transition.
                    A pump tone resonantly drives the $|1\rangle \leftrightarrow |2\rangle$ transition at $\omega_{21} = \omega_{20} - \omega_{10}$.
                } \label{FIG:full_model}
                \end{center}
            \end{figure}

            The proposed quantum transducer consists of a microwave resonator $\hat{a}$ (frequency $\omega_a/2\pi$), a mechanical resonator $\hat{b}$ (frequency $\omega_b/2\pi$), an NV$^0$ center, and an optical resonator $\hat{c}$ (frequency $\omega_c/2\pi$), coupled in series as shown in Fig.~\ref{FIG:full_model}.
            The full-model Hamiltonian is then given by
            \begin{align}
                \label{eq:system_Hamiltonian_SM}
                \begin{split}
                    \hat{H}/\hbar   &=      \omega_{a} \hat{a}^{\dagger} \hat{a} +
                                            \omega_{b} \hat{b}^{\dagger} \hat{b} + 
                                            \omega_{c} \hat{c}^{\dagger} \hat{c} +
                                            \omega_{10}\hat{\sigma}_{11} + \omega_{20}\hat{\sigma}_{22}\\
                                    &\qquad+g_{x} (\hat{a}^{\dagger}\hat{b} + \hat{a}\hat{b}^{\dagger}) +
                                            g_{y} (\hat{b}^{\dagger}\hat{\sigma}_{01} + \hat{b}\hat{\sigma}_{10}) +
                                            g_{02} (\hat{\sigma}_{20}\hat{c} + \hat{\sigma}_{02}\hat{c}^{\dagger}) +
                                            g_{12} (\hat{\sigma}_{21}\hat{c} + \hat{\sigma}_{12}\hat{c}^{\dagger}),
                \end{split}
            \end{align}
            where $\hat{\sigma}_{ij} \equiv |i \rangle\langle j|\,\,(i,j = 0,1,2)$ and $\omega_{ij}$ denotes the transition frequency associated with the transition $|j\rangle \rightarrow |i\rangle$.

            Under this Hamiltonian, the system operator $\hat{O}$ evolves according to the Heisenberg-Langevin equation \cite{GardinerCollett1985}:
            \begin{equation}
                \label{eq:Langevin_equation}
                \frac{d}{dt}\hat{O} = - \frac{i}{\hbar} [ \hat{O}, \hat{H} ]
                                      + \frac{1}{2} \sum_\nu \left( [\hat{L}^{\dagger}_\nu, \hat{O}] \hat{L}_\nu + \hat{L}^{\dagger}_\nu [\hat{O}, \hat{L}_\nu] \right)
                                      + i \sum_\nu \left( [\hat{L}^{\dagger}_\nu, \hat{O}] \hat{L}_{\nu,\mathrm{in}}(t) - \hat{L}^{\dagger}_{\nu,\mathrm{in}}(t) [\hat{O}, \hat{L}_\nu] \right),
            \end{equation}
            where $\hat{L}_\nu$ is the Lindblad operator for the loss channel $\nu$, and $\hat{L}_{\nu,\mathrm{in}}(t)$ is the corresponding input-field operator.
            The index $\nu$ runs over both the population relaxation processes shown in Fig.~\ref{FIG:full_model} and phase relaxation processes.

        \subsubsection{Heisenberg-Langevin equations}
            We introduce the following notation for a system operator,
            \begin{equation}
                \hat{O}_{\vec{\lambda}} = (\hat{a}^{\dagger})^{p} (\hat{a})^{q} (\hat{b}^{\dagger})^{r} (\hat{b})^{s} (\hat{c}^{\dagger})^{u} (\hat{c})^{v} \hat{\sigma}_{mn},
            \end{equation}
            where $\vec{\lambda} \equiv (p,q,r,s,u,v,m,n)$ and $p,q,r,s,u,v$ are non-negative integers.
            In the numerical calculations, we truncate the operator hierarchy by retaining only basis elements with $p,q,r,s,u,v \in \{0,1\}$.
            We verified that the numerical results remain essentially unchanged when higher-order operator products are included.

            A single basis element $\hat{O}_{\vec{\lambda}}$ contains an NV$^0$ operator $\hat{\sigma}_{mn}$.
            Therefore, operators that act as the identity in the NV$^0$ subspace, such as $\hat{a}$, must be represented using the completeness relation $\hat{\sigma}_{00} + \hat{\sigma}_{11} + \hat{\sigma}_{22} = 1$.
            For example, using the completeness relation $\hat{\sigma}_{00} + \hat{\sigma}_{11} + \hat{\sigma}_{22} = 1$, the operator $\hat{a}$ can be written as $\hat{a} = \hat{a}\hat{\sigma}_{00} + \hat{a}\hat{\sigma}_{11} + \hat{a}\hat{\sigma}_{22}$.
            In the shorthand notation, $\hat{O}_{pq,rs,uv,mn}$, this relation is expressed as $\hat{a} = \hat{O}_{01,00,00,00} + \hat{O}_{01,00,00,11} + \hat{O}_{01,00,00,22}$.

            Substituting $\hat{O}_{\vec{\lambda}}$ into Eq.~(\ref{eq:Langevin_equation}), the resulting Heisenberg-Langevin equation can be written in matrix form as
            \begin{equation}
                \label{eq:HL_eq_full_model}
                \frac{d}{dt}\hat{O}_{\vec{\lambda}_1} = \sum_{\vec{\lambda}_2}
                \left( M_{\vec{\lambda}_1,\vec{\lambda}_2}^{(1)}\hat{O}_{\vec{\lambda}_2} + 
                    M_{\vec{\lambda}_1,\vec{\lambda}_2}^{(2)}\hat{a}_{\mathrm{in}}^{\dagger}(t)\hat{O}_{\vec{\lambda}_2} +
                    M_{\vec{\lambda}_1,\vec{\lambda}_2}^{(3)}\hat{O}_{\vec{\lambda}_2}\hat{a}_{\mathrm{in}}(t) + 
                    M_{\vec{\lambda}_1,\vec{\lambda}_2}^{(4)}\hat{c}_{\mathrm{in}}^{\dagger}(t)\hat{O}_{\vec{\lambda}_2} + 
                    M_{\vec{\lambda}_1,\vec{\lambda}_2}^{(5)}\hat{O}_{\vec{\lambda}_2}\hat{c}_{\mathrm{in}}(t)\right).
            \end{equation}
            Here, $\hat{a}_{\mathrm{in}}$ and $\hat{c}_{\mathrm{in}}$ denote the input-field operators of the microwave and optical waveguides, respectively.
            We explicitly retain only these waveguide input fields and neglect the input-field operators associated with relaxation channels other than the waveguide modes.

        \subsubsection{Calculation methods}

            As described in the main text, we calculate the stationary solutions of Eq.~(\ref{eq:HL_eq_full_model}).
            We consider a transduction process in which a single microwave photon is incident from the microwave waveguide, while a coherent pump is incident from the optical waveguide.
            To avoid directly treating the single-photon microwave response, we instead use a coherent-state input from the microwave waveguide \cite{KoshinoPRL2004,KoshinoPRL2007}.
            Since a coherent state is an eigenstate of the input-field operator, the input operators $\hat{a}_{\mathrm{in}}(t)$ and $\hat{c}_{\mathrm{in}}(t)$ are rigorously replaceable with their expectation values, $\langle \hat{a}_{\mathrm{in}}(t) \rangle = E_{\mathrm{s}} e^{-i \mu_{\mathrm{s}} t}$ and $\langle \hat{c}_{\mathrm{in}}(t) \rangle = E_{\mathrm{p}} e^{-i \mu_{\mathrm{p}} t}$.

            Because $\hat{O}_{\vec{\lambda}}$ is driven by the two tones at frequencies $\mu_{\mathrm{p}}$ and $\mu_{\mathrm{s}}$, its stationary solution can be written as
            \begin{equation}
                \langle \hat{O}_{\vec{\lambda}}(t) \rangle  = \sum_{\alpha,\beta} s_{\vec{\lambda}}^{(\alpha,\beta)} e^{-i(\alpha \mu_{\mathrm{p}} + \beta \mu_{\mathrm{s}})t},
            \end{equation}
            where $\alpha$ and $\beta$ are integers. 
            In the present case, the microwave signal is sufficiently weak that it is sufficient to truncate the expansion to the range $-1 \leq \alpha,\beta \leq 1$, as confirmed numerically.
            By taking the expectation value of Eq.~(\ref{eq:HL_eq_full_model}) and substituting this ansatz, we obtain the following set of simultaneous linear equations:
            \begin{align}
                \label{eq:simultaneous_equations_of_FM}
                \sum_{\vec{\lambda}_2} \Bigg[ &\left(M_{\vec{\lambda}_1,\vec{\lambda}_2}^{(1)} + i (\alpha \mu_{\mathrm{p}} + \beta \mu_{\mathrm{s}}) \delta_{\vec{\lambda}_1,\vec{\lambda}_2} \right) s_{\vec{\lambda}_2}^{(\alpha,\beta)}  \nonumber \\
                    &+ E_{\mathrm{s}}\,M_{\vec{\lambda}_1,\vec{\lambda}_2}^{(2)}\,s_{\vec{\lambda}_2}^{(\alpha,\beta-1)}
                    + E_{\mathrm{s}}^{\ast}\,M_{\vec{\lambda}_1,\vec{\lambda}_2}^{(3)}\,s_{\vec{\lambda}_2}^{(\alpha,\beta+1)} \\
                    &+ E_{\mathrm{p}}\,M_{\vec{\lambda}_1,\vec{\lambda}_2}^{(4)}\,s_{\vec{\lambda}_2}^{(\alpha-1,\beta)}
                    + E_{\mathrm{p}}^{\ast}\,M_{\vec{\lambda}_1,\vec{\lambda}_2}^{(5)}\,s_{\vec{\lambda}_2}^{(\alpha+1,\beta)} \Bigg] \nonumber
                = 0,
            \end{align}
            where $\delta$ denotes the Kronecker delta.
            It is of note that, due to the completeness relation $\langle\hat{O}_{00,00,00,00}\rangle + \langle\hat{O}_{00,00,00,11}\rangle + \langle\hat{O}_{00,00,00,22}\rangle = 1$, the above simultaneous equations are not linearly independent.
            We therefore replace the equation for $s_{00,00,00,00}^{(0,0)}$ with the completeness relation.

            Using the optical input-output relation,
            \begin{equation}
                \hat{c}_{\mathrm{out}}(t) = \hat{c}_{\mathrm{in}}(t) - i \sqrt{\kappa_{c}^{e}}\hat{c},
            \end{equation}
            we evaluate the following output photon flux and field amplitudes from $\langle \hat{O}_{\vec{\lambda}}(t) \rangle$, which is obtained by solving Eq.~(\ref{eq:simultaneous_equations_of_FM}), 
            \begin{align}
                \langle \hat{c}_{\mathrm{out}}^{\dagger}(t)\hat{c}_{\mathrm{out}}(t) \rangle
                    &= |E_{\mathrm{p}}|^2 - i \sqrt{\kappa_{c}^{e}} E_{\mathrm{p}}^{\ast} \left( s_{00,00,01,00}^{(1,0)} + s_{00,00,01,11}^{(1,0)} + s_{00,00,01,22}^{(1,0)} \right) \nonumber \\
                    &\qquad\qquad+ i \sqrt{\kappa_{c}^{e}} E_{\mathrm{p}} \left( s_{00,00,10,00}^{(-1,0)} + s_{00,00,10,11}^{(-1,0)} + s_{00,00,10,22}^{(-1,0)} \right) \\
                    &\qquad\qquad+ \kappa_c^e \left( s_{00,00,11,00}^{(0,0)} + s_{00,00,11,11}^{(0,0)} + s_{00,00,11,22}^{(0,0)} \right), \nonumber \\
                \langle \hat{c}_{\mathrm{out}}^{(1,0)}(t)\rangle
                    &= \left[ E_{\mathrm{p}} - i \sqrt{\kappa_{c}^{e}} \left( s_{00,00,01,00}^{(1,0)} + s_{00,00,01,11}^{(1,0)} + s_{00,00,01,22}^{(1,0)} \right) \right] e^{-i \mu_{\mathrm{p}} t}, \\
                \langle \hat{c}_{\mathrm{out}}^{(1,1)}(t)\rangle
                    &= - i \sqrt{\kappa_{c}^{e}} \left( s_{00,00,01,00}^{(1,1)} + s_{00,00,01,11}^{(1,1)} + s_{00,00,01,22}^{(1,1)} \right) e^{-i (\mu_{\mathrm{p}} + \mu_{\mathrm{s}})t},
            \end{align}
            where rapidly oscillating components of $\langle \hat{c}_{\mathrm{out}}^{\dagger}(t)\hat{c}_{\mathrm{out}}(t) \rangle$ are neglected.
            $\langle \hat{c}_{\mathrm{out}}(t)\rangle$ corresponds to the output-field amplitude, and $\langle \hat{c}_{\mathrm{out}}^{(\alpha,\beta)}(t)\rangle$ denotes its component oscillating as $\sim e^{-i(\alpha \mu_{\mathrm{p}} + \beta \mu_{\mathrm{s}})t}$.
            Owing to the large frequency separation between the target frequency $\mu_{\mathrm{p}} + \mu_{\mathrm{s}}$ and the pump frequency $\mu_{\mathrm{p}}$, the unwanted component at $\mu_{\mathrm{p}}$ can be removed using a frequency filter.
            Therefore, we define the total output photon flux $R_{\mathrm{tot}}$ and the target coherent photon flux $R_{\mathrm{coh}}$ as
            \begin{align}
                R_{\mathrm{tot}} &= \langle \hat{c}_{\mathrm{out}}^{\dagger}(t)\hat{c}_{\mathrm{out}}(t) \rangle  - \left| \langle \hat{c}_{\mathrm{out}}^{(1,0)}(t)\rangle \right|^2, \\
                R_{\mathrm{coh}} &= \left| \langle \hat{c}_{\mathrm{out}}^{(1,1)}(t)\rangle \right|^2.
            \end{align}
            Here, $\left| \langle \hat{c}_{\mathrm{out}}^{(1,0)}(t)\rangle \right|^2$ represents the reflected pump photon flux. 
            We numerically confirm that $|\langle \hat{c}_{\mathrm{out}}^{(1,-1)}(t)\rangle|^2$ and higher-order sidebands are small enough to be neglected.

            To calculate the single-photon microwave response, we use the correspondence between coherent and number states as mentioned in the main text.
            Finally, we obtain the total photon-detection probability $P_{\mathrm{tot}}$ and the coherent conversion efficiency of the transduction $P_{\mathrm{coh}}$ as
            \begin{align}
                P_{\mathrm{tot}} &= \lim_{\left| E_{\mathrm{s}} \right|^2 \rightarrow 0} \frac{dR_{\mathrm{tot}}}{d\left| E_{\mathrm{s}} \right|^2}, \\
                P_{\mathrm{coh}} &= \lim_{\left| E_{\mathrm{s}} \right|^2 \rightarrow 0} \frac{dR_{\mathrm{coh}}}{d\left| E_{\mathrm{s}} \right|^2}.
            \end{align}
            Their difference, $P_{\mathrm{tot}} - P_{\mathrm{coh}}$, represents the detection probability $P_{\mathrm{inc}}$ of incoherent photons generated by spontaneous emission after dephasing at the NV$^0$ excited state $|2\rangle$.

            Even in the absence of a signal, the pump weakly excites the NV$^0$ and yields a dark count rate.
            The dark count rate is given by
            \begin{equation}
                R_{\mathrm{d}} = \lim_{|E_{\mathrm{s}}|^2 \rightarrow 0} R_{\mathrm{tot}}.
            \end{equation}

            As discussed in Sec.~\ref{sec:entanglement_generation}, the phase shift induced by the transducer plays an important role in remote entanglement generation using a single-click scheme.
            Here, we focus on the transducer-induced phase difference $\Delta\theta$ between the two interference paths.
            It is given by
            \begin{equation}
                \label{eq:phase_difference}
                \Delta\theta = \theta_{\mathrm{A}} - \theta_{\mathrm{B}},
            \end{equation}
            where $\theta_j$ denotes the phase imparted by the transducer at node $j \in \{\mathrm{A},\mathrm{B}\}$ and is defined as
            \begin{equation}
                \theta_j = \arg\left[\lim_{E_{\mathrm{s}} \rightarrow 0} \frac{d}{d E_{\mathrm{s}}} \langle \hat{c}_{\mathrm{out}}^{(1,1)}(t)\rangle_j\right].
            \end{equation}
            
    \subsection{Reduced model} \label{subsec:reduced_model}

\begin{figure}[htbp]
                \begin{center}
                \includegraphics[width=0.5\columnwidth]{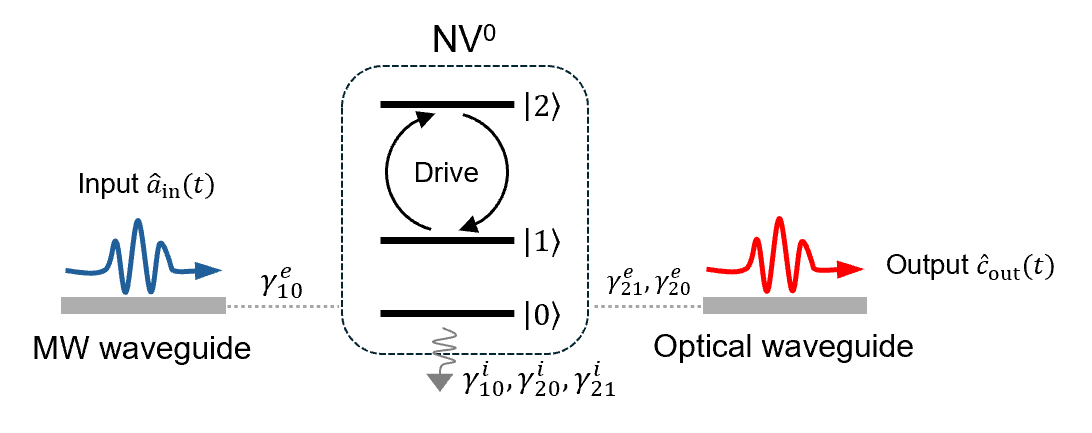}
                \caption{
                    Schematic illustration of the transduction process in the reduced model.
                    The microwave and optical waveguides are assumed to couple directly to an NV$^0$ center.
                    $\gamma_{10}^e\,(\gamma_{10}^i)$ is the external coupling (internal loss) rate enhanced by microwave and mechanical cavities,
                    while $\gamma_{20}^e\,(\gamma_{20}^i), \gamma_{21}^e\,(\gamma_{21}^i)$ denote the corresponding rates enhanced by the optical cavity.
                } \label{FIG:reduced_model}
                \end{center}
            \end{figure}

        From the full model, we can derive a reduced model [Fig.~\ref{FIG:reduced_model}] in which the NV$^0$ center is directly coupled to the microwave and optical waveguides.
        This model enables us to estimate the conversion efficiencies with significantly lower computational cost while retaining quantitative accuracy (see Fig.~\ref{FIG:numerical_results}).

        \subsubsection{Derivation of the reduced model}

            The reduced model is obtained by adiabatically eliminating the three cavity modes, $\hat{a}$, $\hat{b}$, and $\hat{c}$, under the steady-state approximation.
            In the rotating frame of the respective modes, their Heisenberg-Langevin equations are given by
            \begin{align}
                \frac{d}{dt} \hat{a} &=     - \frac{\kappa_a^t}{2} \hat{a}
                                            - i g_{x} \hat{b}
                                            - i \sqrt{\kappa_a^e} \hat{a}_{\mathrm{in}}^{e}(t)
                                            - i \sqrt{\kappa_a^i} \hat{a}_{\mathrm{in}}^{i}(t), \label{eq:ss_approximation_start} \\
                \frac{d}{dt} \hat{b} &=     - \frac{\kappa_b^i}{2} \hat{b}
                                            - i g_{x} \hat{a}
                                            - i g_{y} \hat{\sigma}_{01}
                                            - i \sqrt{\kappa_b^i} \hat{b}_{\mathrm{in}}^{i}(t), \\
                \frac{d}{dt} \hat{c} &=     - \frac{\kappa_c^t}{2} \hat{c}
                                            - i g_{02} \hat{\sigma}_{02}
                                            - i g_{12} \hat{\sigma}_{12}
                                            - i \sqrt{\kappa_c^e} \hat{c}_{\mathrm{in}}^{e}(t)
                                            - i \sqrt{\kappa_c^i} \hat{c}_{\mathrm{in}}^{i}(t), \label{eq:ss_approximation_end}
            \end{align}
            where $\kappa_a^t = \kappa_a^e + \kappa_a^i$ and $\kappa_c^t = \kappa_c^e + \kappa_c^i$.
            $\hat{a}_{\mathrm{in}}^{e}(t)$ and $\hat{c}_{\mathrm{in}}^{e}(t)$ denote the input-field operators of the microwave and optical waveguides, respectively.
            The operators $\hat{a}_{\mathrm{in}}^{i}(t)$, $\hat{b}_{\mathrm{in}}^{i}(t)$ and $\hat{c}_{\mathrm{in}}^{i}(t)$ denote the corresponding environmental input fields responsible for intrinsic losses.
            
            To derive the reduced model, we first eliminate the microwave and optical cavity modes, $\hat{a}$ and $\hat{c}$, and then eliminate the mechanical mode $\hat{b}$.
            Substituting the resulting expressions into the input-output relations
            \begin{align}
                \hat{a}_{\mathrm{out}}^{e/i}(t) &= \hat{a}_{\mathrm{in}}^{e/i}(t) - i\sqrt{\kappa_a^{e/i}} \hat{a}, \\
                \hat{b}_{\mathrm{out}}^{i}(t)   &= \hat{b}_{\mathrm{in}}^{i}(t)   - i\sqrt{\kappa_b^{i}}   \hat{b}, \\
                \hat{c}_{\mathrm{out}}^{e/i}(t) &= \hat{c}_{\mathrm{in}}^{e/i}(t) - i\sqrt{\kappa_c^{e/i}} \hat{c},
            \end{align}
            we obtain the following effective input-output relations:
            \begin{align}
                \hat{a}_{\mathrm{out}}^{e}(t) &= \left( 1 - \frac{2 \kappa_a^e \kappa_b^i}{\kappa_a^t \kappa_b^t} \right) \hat{a}_{\mathrm{in}}^{e}(t)
                                                    - \frac{2 \kappa_b^i \sqrt{\kappa_a^e \kappa_a^i}}{\kappa_a^t \kappa_b^t} \hat{a}_{\mathrm{in}}^{i}(t)
                                                    + i \frac{4 g_{x} \sqrt{\kappa_a^e \kappa_b^i}}{\kappa_a^t \kappa_b^t} \hat{b}_{\mathrm{in}}^{i}(t)
                                                    + i \frac{4 g_{x} g_{y} \sqrt{\kappa_a^e}}{\kappa_a^t \kappa_b^t} \hat{\sigma}_{01}, \\
                \hat{a}_{\mathrm{out}}^{i}(t) &= \left( 1 - \frac{2 \kappa_a^i \kappa_b^i}{\kappa_a^t \kappa_b^t} \right) \hat{a}_{\mathrm{in}}^{i}(t)
                                                    - \frac{2 \kappa_b^i \sqrt{\kappa_a^e \kappa_a^i}}{\kappa_a^t \kappa_b^t} \hat{a}_{\mathrm{in}}^{e}(t)
                                                    + i \frac{4 g_{x} \sqrt{\kappa_a^i \kappa_b^i}}{\kappa_a^t \kappa_b^t} \hat{b}_{\mathrm{in}}^{i}(t)
                                                    + i \frac{4 g_{x} g_{y} \sqrt{\kappa_a^i}}{\kappa_a^t \kappa_b^t} \hat{\sigma}_{01}, \\
                \hat{b}_{\mathrm{out}}^{i}(t) &= \left( 1 - \frac{2 \kappa_b^i}{\kappa_b^t} \right) \hat{b}_{\mathrm{in}}^{i}(t)
                                                    + i \frac{4 g_{x} \sqrt{\kappa_a^e \kappa_b^i}}{\kappa_a^t \kappa_b^t} \hat{a}_{\mathrm{in}}^{e}(t)
                                                    + i \frac{4 g_{x} \sqrt{\kappa_a^i \kappa_b^i}}{\kappa_a^t \kappa_b^t} \hat{a}_{\mathrm{in}}^{i}(t)
                                                    - \frac{2 g_{y} \sqrt{\kappa_b^i}}{\kappa_b^t} \hat{\sigma}_{01}, \\
                \hat{c}_{\mathrm{out}}^{e}(t) &= \left( 1 - \frac{2 \kappa_c^e}{\kappa_c^t} \right) \hat{c}_{\mathrm{in}}^{e}(t)
                                                    - \frac{2 \sqrt{\kappa_c^e \kappa_c^i}}{\kappa_c^t} \hat{c}_{\mathrm{in}}^{i}(t)
                                                    - \frac{2 g_{02} \sqrt{\kappa_c^e}}{\kappa_c^t} \hat{\sigma}_{02}
                                                    - \frac{2 g_{12} \sqrt{\kappa_c^e}}{\kappa_c^t} \hat{\sigma}_{12}, \\
                \hat{c}_{\mathrm{out}}^{i}(t) &= \left( 1 - \frac{2 \kappa_c^i}{\kappa_c^t} \right) \hat{c}_{\mathrm{in}}^{i}(t)
                                                    - \frac{2 \sqrt{\kappa_c^e \kappa_c^i}}{\kappa_c^t} \hat{c}_{\mathrm{in}}^{e}(t)
                                                    - \frac{2 g_{02} \sqrt{\kappa_c^i}}{\kappa_c^t} \hat{\sigma}_{02}
                                                    - \frac{2 g_{12} \sqrt{\kappa_c^i}}{\kappa_c^t} \hat{\sigma}_{12}.
            \end{align}
            By regarding these equations as effective input-output relations between the NV$^0$ and the corresponding environmental modes, we obtain the following reduced parameters:
            \begin{align}
                \gamma_{20}^{e} &= \frac{4 g_{02}^{2} \kappa_c^e}{(\kappa_c^t)^2}, \label{eq:reduced_param_start} \\
                \gamma_{20}^{i} &= \gamma_{20} + \frac{4 g_{02}^{2} \kappa_c^i}{(\kappa_c^t)^2}, \\
                \gamma_{21}^{e} &= \frac{4 g_{12}^{2} \kappa_c^e}{(\kappa_c^t)^2}, \\
                \gamma_{21}^{i} &= \gamma_{21} + \frac{4 g_{12}^{2} \kappa_c^i}{(\kappa_c^t)^2}, \\
                \kappa_{b}^{t}  &= \kappa_{b}^{i} + \frac{4 g_{x}^{2}}{\kappa_a^t}, \\
                \gamma_{10}^{e} &= \frac{16 g_{x}^{2} g_{y}^{2} \kappa_a^e}{(\kappa_a^t)^2 (\kappa_b^t)^2}, \\
                \gamma_{10}^{i} &= \gamma_{10} + \frac{4 g_{y}^{2} \kappa_b^i}{(\kappa_b^t)^2} + \frac{16 g_{x}^{2} g_{y}^{2} \kappa_a^i}{(\kappa_a^t)^2 (\kappa_b^t)^2}. \label{eq:reduced_param_end} 
            \end{align}
            Here, we define the total decay rates as $\gamma_{10}^{t} = \gamma_{10}^{e} + \gamma_{10}^{i}$, $\gamma_{20}^{t} = \gamma_{20}^{e} + \gamma_{20}^{i}$ and $\gamma_{21}^{t} = \gamma_{21}^{e} + \gamma_{21}^{i}$.
            The resulting parameter values for the present settings are listed in Table~\ref{tab:reduced_parameters}.

            \begin{table}[b]
                \centering
                \begin{minipage}{0.38\textwidth}
                    \caption{Reduced parameters derived from the parameter values listed in Table~\ref{tab:parameters}.} \label{tab:reduced_parameters}
                    \centering
                    \begin{ruledtabular}
                        \begin{tabular}{lrc}
                            \textrm{Parameter} & \textrm{Value}\\
                            \hline
                            
                            \multicolumn{2}{l}{\textit{Internal decay rates}}\\
                            \textrm{$|1\rangle\to|0\rangle$ decay rate, $\gamma_{10}^i/2\pi$}                    & $0.81~\mathrm{MHz}$ \\
                            \textrm{$|2\rangle\to|0\rangle$ decay rate, $\gamma_{20}^i/2\pi$}                    & $7.5~\mathrm{MHz}$ \\
                            \textrm{$|2\rangle\to|1\rangle$ decay rate, $\gamma_{21}^i/2\pi$}                    & $3.7~\mathrm{MHz}$ \\
                            \hline
                            
                            \multicolumn{2}{l}{\textit{External coupling rates}}\\
                            \textrm{$|0\rangle\leftrightarrow|1\rangle$ coupling rate, $\gamma_{10}^e/2\pi$}     & $1.2~\mathrm{MHz}$ \\
                            \textrm{$|0\rangle\leftrightarrow|2\rangle$ coupling rate, $\gamma_{20}^e/2\pi$}     & $14~\mathrm{MHz}$ \\
                            \textrm{$|1\rangle\leftrightarrow|2\rangle$ coupling rate, $\gamma_{21}^e/2\pi$}     & $0.55~\mathrm{MHz}$ \\
                        \end{tabular}
                    \end{ruledtabular}
                \end{minipage}
            \end{table}

        \subsubsection{Heisenberg-Langevin equations}
            Employing the reduced parameters in Eqs.~(\ref{eq:reduced_param_start})-(\ref{eq:reduced_param_end}), the Heisenberg-Langevin equations for the reduced model are then given by
            \begin{align}
                &\frac{d}{dt} \hat{\sigma}_{00} = \gamma_{20}^{t} \hat{\sigma}_{22} + \gamma_{10}^{t} \hat{\sigma}_{11}
                                                    - i \sqrt{\gamma_{20}^{e}} \hat{c}_{\mathrm{in}}^{\dagger}(t) \hat{\sigma}_{02}
                                                    + i \sqrt{\gamma_{20}^{e}} \hat{\sigma}_{20} \hat{c}_{\mathrm{in}}(t) 
                                                    - i \sqrt{\gamma_{10}^{e}} \hat{a}_{\mathrm{in}}^{\dagger}(t) \hat{\sigma}_{01}
                                                    + i \sqrt{\gamma_{10}^{e}} \hat{\sigma}_{10} \hat{a}_{\mathrm{in}}(t), \label{eq:reduced_equations_start} \\
                &\frac{d}{dt} \hat{\sigma}_{11} = \gamma_{21}^{t} \hat{\sigma}_{22} - \gamma_{10}^{t} \hat{\sigma}_{11}
                                                    - i \sqrt{\gamma_{21}^{e}} \hat{c}_{\mathrm{in}}^{\dagger}(t) \hat{\sigma}_{12}
                                                    + i \sqrt{\gamma_{21}^{e}} \hat{\sigma}_{21} \hat{c}_{\mathrm{in}}(t) 
                                                    + i \sqrt{\gamma_{10}^{e}} \hat{a}_{\mathrm{in}}^{\dagger}(t) \hat{\sigma}_{01}
                                                    - i \sqrt{\gamma_{10}^{e}} \hat{\sigma}_{10} \hat{a}_{\mathrm{in}}(t), \\
                &\frac{d}{dt} \hat{\sigma}_{22} =  - (\gamma_{20}^{t} + \gamma_{21}^{t}) \hat{\sigma}_{22}
                                                    + i \hat{c}_{\mathrm{in}}^{\dagger}(t) (\sqrt{\gamma_{20}^{e}} \hat{\sigma}_{02} + \sqrt{\gamma_{21}^{e}} \hat{\sigma}_{12})
                                                    - i (\sqrt{\gamma_{20}^{e}} \hat{\sigma}_{20} + \sqrt{\gamma_{21}^{e}} \hat{\sigma}_{21}) \hat{c}_{\mathrm{in}}(t), \\
                &\frac{d}{dt} \hat{\sigma}_{01} = (- i \omega_{10} - \gamma_{10}^t/2 - \gamma_{10}^{\phi}) \hat{\sigma}_{01}
                                                    - i \sqrt{\gamma_{21}^{e}} \hat{c}_{\mathrm{in}}^{\dagger}(t) \hat{\sigma}_{02}
                                                    + i \sqrt{\gamma_{20}^{e}} \hat{\sigma}_{21} \hat{c}_{\mathrm{in}}(t) 
                                                    - i \sqrt{\gamma_{10}^{e}} (\hat{\sigma}_{00} - \hat{\sigma}_{11}) \hat{a}_{\mathrm{in}}(t), \\
                &\frac{d}{dt} \hat{\sigma}_{02} = (- i \omega_{20} - \gamma_{20}^t/2 - \gamma_{21}^t/2 - \gamma_{20}^{\phi}) \hat{\sigma}_{02}
                                                    - i ( \sqrt{\gamma_{20}^{e}}\hat{\sigma}_{00} + \sqrt{\gamma_{21}^{e}}\hat{\sigma}_{01} - \sqrt{\gamma_{20}^{e}}\hat{\sigma}_{22}) \hat{c}_{\mathrm{in}}(t)
                                                    + i \sqrt{\gamma_{10}^{e}} \hat{\sigma}_{12} \hat{a}_{\mathrm{in}}(t), \\
                &\frac{d}{dt} \hat{\sigma}_{12} = (- i \omega_{21} - \gamma_{10}^t/2 - \gamma_{20}^t/2 - \gamma_{21}^t/2 - \gamma_{21}^{\phi}) \hat{\sigma}_{12}
                                                    - i ( \sqrt{\gamma_{20}^{e}}\hat{\sigma}_{10} + \sqrt{\gamma_{21}^{e}}\hat{\sigma}_{11} - \sqrt{\gamma_{21}^{e}}\hat{\sigma}_{22}) \hat{c}_{\mathrm{in}}(t)
                                                    + i \sqrt{\gamma_{10}^{e}} \hat{a}_{\mathrm{in}}^{\dagger}(t) \hat{\sigma}_{02}, \label{eq:reduced_equations_end}
            \end{align}
            where $\gamma_{10}^{\phi} = \gamma_{1}^{\phi}$, $\gamma_{20}^{\phi} = \gamma_{2}^{\phi}$ and $\gamma_{21}^{\phi} = \gamma_{1}^{\phi} + \gamma_{2}^{\phi}$.
            $\gamma_{1}^{\phi}$ and $\gamma_{2}^{\phi}$ are the pure-dephasing rates of levels $|1\rangle$ and $|2\rangle$, respectively.
            $\hat{a}_{\mathrm{in}}$ and $\hat{c}_{\mathrm{in}}$ denote the input-field operators of the microwave and optical waveguides in the reduced model, respectively.
            The expressions for $\hat{\sigma}_{10}$, $\hat{\sigma}_{20}$, and $\hat{\sigma}_{21}$ are obtained by taking the Hermitian conjugates of those for $\hat{\sigma}_{01}$, $\hat{\sigma}_{02}$, and $\hat{\sigma}_{12}$, respectively.

            We rewrite these equations in matrix form as
            \begin{equation}
                \label{eq:HL_eq_reduced_model}
                \frac{d}{dt}\hat{\sigma}_{ij} = \sum_{m,n} \left( \tilde{M}_{ij,mn}^{(1)}\hat{\sigma}_{mn} + \tilde{M}_{ij,mn}^{(2)}\hat{a}_{\mathrm{in}}^{\dagger}(t)\hat{\sigma}_{mn} + \tilde{M}_{ij,mn}^{(3)}\hat{\sigma}_{mn}\hat{a}_{\mathrm{in}}(t) + \tilde{M}_{ij,mn}^{(4)}\hat{c}_{\mathrm{in}}^{\dagger}(t)\hat{\sigma}_{mn} + \tilde{M}_{ij,mn}^{(5)}\hat{\sigma}_{mn}\hat{c}_{\mathrm{in}}(t)\right).
            \end{equation}

        \subsubsection{Calculation methods}
            As in the full model, we calculate the stationary solutions of Eq.~(\ref{eq:HL_eq_reduced_model}) for coherent-state inputs.

            The stationary solution of Eq.~(\ref{eq:HL_eq_reduced_model}) is given by
            \begin{equation}
                \langle \hat{\sigma}_{mn}(t) \rangle  = \sum_{\alpha,\beta} s_{mn}^{(\alpha,\beta)} e^{-i(\alpha \mu_{\mathrm{p}} + \beta \mu_{\mathrm{s}})t},
            \end{equation}
            where $\alpha$ and $\beta$ are integers. 
            By taking the expectation value of Eq.~(\ref{eq:HL_eq_reduced_model}) and substituting this ansatz, we obtain the following set of simultaneous linear equations:
            \begin{align}
                \label{eq:simultaneous_equations_of_RM}
                \sum_{m_2,n_2} \Bigg[ &\left(\tilde{M}_{m_1n_1,m_2n_2}^{(1)} + i (\alpha \mu_{\mathrm{p}} + \beta \mu_{\mathrm{s}}) \delta_{m_1m_2} \delta_{n_1n_2} \right) s_{m_2n_2}^{(\alpha,\beta)}  \nonumber \\
                    &+ E_{\mathrm{s}}\,\tilde{M}_{m_1n_1,m_2n_2}^{(2)}\,s_{m_2n_2}^{(\alpha,\beta-1)}
                    + E_{\mathrm{s}}^{\ast}\,\tilde{M}_{m_1n_1,m_2n_2}^{(3)}\,s_{m_2n_2}^{(\alpha,\beta+1)} \\
                    &+ E_{\mathrm{p}}\,\tilde{M}_{m_1n_1,m_2n_2}^{(4)}\,s_{m_2n_2}^{(\alpha-1,\beta)}
                    + E_{\mathrm{p}}^{\ast}\,\tilde{M}_{m_1n_1,m_2n_2}^{(5)}\,s_{m_2n_2}^{(\alpha+1,\beta)} \Bigg] \nonumber
                = 0,
            \end{align}
            where $\delta$ denotes the Kronecker delta.
            Because of the completeness relation $\langle\hat{\sigma}_{00}\rangle + \langle\hat{\sigma}_{11}\rangle + \langle\hat{\sigma}_{22}\rangle = 1$, the equation for $s_{00}^{(0,0)}$ is replaced by $s_{00}^{(0,0)} + s_{11}^{(0,0)} + s_{22}^{(0,0)} = 1$.

            Using the optical input-output relation,
            \begin{equation}
                \hat{c}_{\mathrm{out}}(t) = \hat{c}_{\mathrm{in}}(t) - i \sqrt{\gamma_{21}^{e}}\hat{\sigma}_{12} - i \sqrt{\gamma_{20}^{e}}\hat{\sigma}_{02},
            \end{equation}
            we evaluate the following output photon flux and amplitudes from $\langle \hat{\sigma}_{mn}(t) \rangle$, which is obtained by solving Eq.~(\ref{eq:simultaneous_equations_of_RM}), 
            \begin{align}
                \langle \hat{c}_{\mathrm{out}}^{\dagger}(t)\hat{c}_{\mathrm{out}}(t) \rangle
                    &= |E_{\mathrm{p}}|^2 - i \sqrt{\gamma_{21}^{e}} E_{\mathrm{p}}^{\ast} s_{12}^{(1,0)} - i \sqrt{\gamma_{20}^{e}} E_{\mathrm{p}}^{\ast} s_{02}^{(1,0)}
                    + i \sqrt{\gamma_{21}^{e}} E_{\mathrm{p}} s_{21}^{(-1,0)} + i \sqrt{\gamma_{20}^{e}} E_{\mathrm{p}} s_{20}^{(-1,0)}
                    + (\gamma_{21}^{e} + \gamma_{20}^{e}) s_{22}^{(0,0)}, \\
                \langle \hat{c}_{\mathrm{out}}^{(1,0)}(t)\rangle
                    &= \left[ E_{\mathrm{p}} - i \sqrt{\gamma_{21}^{e}} s_{12}^{(1,0)} - i \sqrt{\gamma_{20}^{e}} s_{02}^{(1,0)} \right] e^{-i \mu_{\mathrm{p}} t}, \\
                \langle \hat{c}_{\mathrm{out}}^{(1,1)}(t)\rangle
                    &= \left[ - i \sqrt{\gamma_{21}^{e}} s_{12}^{(1,1)} - i \sqrt{\gamma_{20}^{e}} s_{02}^{(1,1)} \right] e^{-i (\mu_{\mathrm{p}} + \mu_{\mathrm{s}})t} \label{eq:reference_last},
            \end{align}
            where rapidly oscillating components of $\langle \hat{c}_{\mathrm{out}}^{\dagger}(t)\hat{c}_{\mathrm{out}}(t) \rangle$ are neglected.

            Using Eqs.~(\ref{eq:HL_eq_reduced_model})--(\ref{eq:reference_last}), one can obtain the calculation results following the same procedure as for the full model.

    \subsection{Phenomenological model} \label{subsec:phenomenological_model}

        A standard approach to evaluating the performance of quantum transducers is to analyze their frequency response.
        To this end, the transduction efficiency and bandwidth can be conveniently characterized in terms of cooperativities.
        However, this approach is applicable only to linear transduction processes.

        Here, we introduce the phenomenological model of our transducer [Fig.~\ref{FIG:phenomenological_model}], in which the transition operators are approximated by bosonic operators in the weak-excitation regime.
        This approximation enables us to derive an analytical expression for the frequency-dependent transduction efficiency.

        \subsubsection{Derivation of the phenomenological Hamiltonian}

            \begin{figure}[htbp]
                \begin{center}
                \includegraphics[width=\columnwidth]{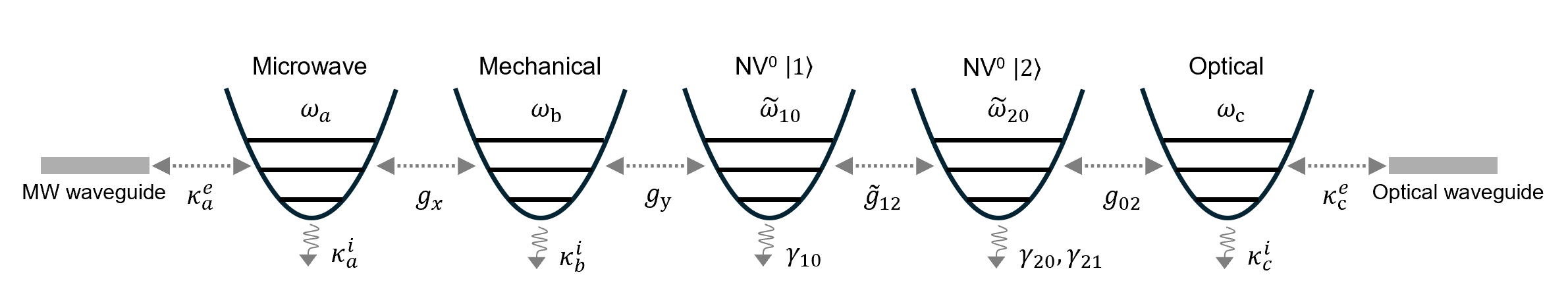}
                \caption{
                    Phenomenological model for the full model.
                } \label{FIG:phenomenological_model}
                \end{center}
            \end{figure}

            We consider the system Hamiltonian in Eq.~(\ref{eq:system_Hamiltonian_SM}) under the pump field with amplitude $E_{\mathrm{p}}$, which is incorporated through the expectation value of the optical cavity mode amplitude.
            In the rotating frame at the pump frequency $\mu_{\mathrm{p}}$, the Hamiltonian is given by
            \begin{align}
                \begin{split}
                    \hat{H}/\hbar   &=      \omega_{a} \hat{a}^{\dagger} \hat{a} +
                                            \omega_{b} \hat{b}^{\dagger} \hat{b} + 
                                            (\omega_{c} - \mu_{\mathrm{p}}) \hat{c}^{\dagger} \hat{c} +
                                            \omega_{10}\hat{\sigma}_{11} + (\omega_{20} - \mu_{\mathrm{p}})\hat{\sigma}_{22} \\
                                    &\quad+g_{x} (\hat{a}^{\dagger}\hat{b} + \hat{a}\hat{b}^{\dagger}) +
                                            g_{y} (\hat{b}^{\dagger}\hat{\sigma}_{01} + \hat{b}\hat{\sigma}_{10}) +
                                            g_{02} (\hat{\sigma}_{20}\hat{c} + \hat{\sigma}_{02}\hat{c}^{\dagger}) +
                                            g_{12} (\hat{\sigma}_{21}\hat{c} + \hat{\sigma}_{12}\hat{c}^{\dagger}).
                \end{split}
            \end{align}
            We decompose the optical cavity operator as $\hat{c} = \langle \hat{c} \rangle + \delta\hat{c},$ where $\langle \hat{c} \rangle$ is approximated by the expectation value of the bare optical cavity mode amplitude driven by the pump field:
            \begin{equation}
                \langle \hat{c} \rangle = \frac{-i \sqrt{\kappa_c^e} E_{\mathrm{p}}}{(\kappa_c^e + \kappa_c^i)/2 -i (\mu_{\mathrm{p}} - \omega_c)}.
            \end{equation}
            Here, $\delta\hat{c}$ denotes the displaced cavity operator.

            We first include the pump-induced coupling $g_{02} (\langle\hat{c}\rangle\hat{\sigma}_{20} + \langle\hat{c}\rangle^{\ast}\hat{\sigma}_{02})$ in the unperturbed Hamiltonian by diagonalizing the $|0\rangle$-$|2\rangle$ subspace.
            The resulting dressed energies are
            \begin{align}
                \tilde{\omega}_{0} = \frac{\omega_{20} - \mu_{\mathrm{p}}}{2} - \sqrt{\left( \frac{\omega_{20} - \mu_{\mathrm{p}}}{2} \right)^2 + |\langle\hat{c}\rangle|^2 g_{02}^2}, \\
                \tilde{\omega}_{2} = \frac{\omega_{20} - \mu_{\mathrm{p}}}{2} + \sqrt{\left( \frac{\omega_{20} - \mu_{\mathrm{p}}}{2} \right)^2 + |\langle\hat{c}\rangle|^2 g_{02}^2}.
            \end{align}
            The pump field also induces an effective coupling between states $|1\rangle$ and $|2\rangle$, $\tilde{g}_{12}=\langle\hat{c}\rangle g_{12}$.
            In contrast, we neglect the fluctuation-mediated term $g_{12} (\hat{\sigma}_{21}\delta\hat{c} + \hat{\sigma}_{12}\delta\hat{c}^{\dagger})$.

            Finally, since we investigate the single-photon response (accordingly, the linear response) of this setup, we bosonize the relevant transitions by replacing $\hat{\sigma}_{01}$ and $\hat{\sigma}_{02}$ with bosonic operators $\hat{d}_1$ and $\hat{d}_2$, respectively, which satisfy $[\hat{d}_i, \hat{d}^{\dagger}_j] = \delta_{ij}\,(i,j=1,2)$.
            This yields the following phenomenological Hamiltonian, illustrated in Fig.~\ref{FIG:phenomenological_model}:
            \begin{align}
                \begin{split}
                    \hat{H}/\hbar   &=      \omega_{a} \hat{a}^{\dagger} \hat{a} +
                                            \omega_{b} \hat{b}^{\dagger} \hat{b} + 
                                            (\omega_{c} - \mu_{\mathrm{p}}) \hat{c}^{\dagger}\hat{c} +
                                            \tilde{\omega}_{10}\hat{d}^{\dagger}_1\hat{d}_1 + \tilde{\omega}_{20}\hat{d}^{\dagger}_2\hat{d}_2 \\
                                    &\qquad+g_{x} (\hat{a}^{\dagger}\hat{b} + \hat{a}\hat{b}^{\dagger}) +
                                            g_{y} (\hat{b}^{\dagger}\hat{d}_1 + \hat{b}\hat{d}^{\dagger}_1) +
                                            (\tilde{g}_{12}^{\ast} \hat{d}^{\dagger}_1\hat{d}_2 + \tilde{g}_{12} \hat{d}_1\hat{d}^{\dagger}_2) +
                                            g_{02} (\hat{d}^{\dagger}_2\hat{c} + \hat{d}_2\hat{c}^{\dagger}),
                \end{split}
            \end{align}
            where $\tilde{\omega}_{20} = \tilde{\omega}_2 - \tilde{\omega}_0$, $\tilde{\omega}_{10} = \omega_{10} - \tilde{\omega}_0$ and $\tilde{g}_{12} = \langle\hat{c}\rangle g_{12}$.
            For notational simplicity, we relabel the fluctuation operator $\delta\hat{c}$ as $\hat{c}$.

        \subsubsection{Heisenberg-Langevin equations}
            The Heisenberg-Langevin equations for the phenomenological model are given by 
            \begin{align}
                &\frac{d}{dt}\hat{a}
                    = \left(- i \omega_a - \frac{\kappa_a^t}{2}\right) \hat{a}
                    - i g_{x} \hat{b} 
                    - i \sqrt{\kappa_a^e} \hat{a}_{\mathrm{in}}(t), \\
                &\frac{d}{dt}\hat{b}
                    = \left(- i \omega_b - \frac{\kappa_b^i}{2}\right) \hat{b}
                    - i g_{x} \hat{a}
                    - i g_{y} \hat{d}_{1}, \\
                &\frac{d}{dt}\hat{d}_{1}
                    = \left(- i \tilde{\omega}_{10} - \frac{\gamma_{10}}{2}\right) \hat{d}_{1}
                    - i g_{y} \hat{b}
                    - i \tilde{g}_{12}^{\ast} \hat{d}_{2}, \\
                &\frac{d}{dt}\hat{d}_{2}
                    = \left(- i \tilde{\omega}_{20} - \frac{\gamma_{20} + \gamma_{21}}{2}\right) \hat{d}_{2}
                    - i \tilde{g}_{12} \hat{d}_{1}
                    - i g_{02} \hat{c}, \\
                &\frac{d}{dt}\hat{c}
                    = \left(- i (\omega_c - \mu_{\mathrm{p}}) - \frac{\kappa_c^t}{2}\right) \hat{c}
                    - i g_{02} \hat{d}_2
                    - i \sqrt{\kappa_c^e} \hat{c}_{\mathrm{in}}(t),
            \end{align}
            where $\hat{a}_{\mathrm{in}}(t)$ and $\hat{c}_{\mathrm{in}}(t)$ are the input-field operators from the microwave and optical waveguides, respectively.
            These operators satisfy the input-output relations
            \begin{align}
                \hat{a}_{\mathrm{out}}(t) = \hat{a}_{\mathrm{in}}(t) - i\sqrt{\kappa_a^e}\hat{a}, \\
                \hat{c}_{\mathrm{out}}(t) = \hat{c}_{\mathrm{in}}(t) - i\sqrt{\kappa_c^e}\hat{c}.
            \end{align}

            These equations can be written in matrix form as
            \begin{align}
                \frac{d}{dt}\hat{\mathbf{a}}(t) = \mathbf{A} \hat{\mathbf{a}}(t) + \mathbf{B} \hat{\mathbf{a}}_{\mathrm{in}}(t), \\
                \hat{\mathbf{a}}_{\mathrm{out}}(t) = \mathbf{C} \hat{\mathbf{a}}(t) + \mathbf{D} \hat{\mathbf{a}}_{\mathrm{in}}(t),
            \end{align}
            where $\hat{\mathbf{a}}(t) = (\hat{a}(t), \hat{b}(t), \hat{d}_1(t), \hat{d}_2(t), \hat{c}(t))^{T}$, $\hat{\mathbf{a}}_{\mathrm{in}}(t) = (\hat{a}_{\mathrm{in}}(t), \hat{c}_{\mathrm{in}}(t))^{T}$ and $\hat{\mathbf{a}}_{\mathrm{out}}(t) = (\hat{a}_{\mathrm{out}}(t), \hat{c}_{\mathrm{out}}(t))^{T}$.
            The matrices are given by 
            \begin{align}
                &\mathbf{A} =
                    \begin{pmatrix}
                    - i\omega_{a} - \frac{\kappa_a^t}{2} & - i g_{x} & 0 & 0 & 0 \\
                    - i g_{x} & - i\omega_b - \frac{\kappa_b^i}{2} & - i g_{y} & 0 & 0 \\
                    0 & - i g_{y} & - i\tilde{\omega}_{10} - \frac{\gamma_{10}}{2} & - i \tilde{g}_{12}^{\ast} & 0 \\
                    0 & 0 & - i \tilde{g}_{12} & - i\tilde{\omega}_{20} -  \frac{\gamma_{20} + \gamma_{21}}{2} & -i g_{02} \\
                    0 & 0 & 0 & -i g_{02} & - i(\omega_{c} - \mu_{\mathrm{p}}) - \frac{\kappa_c^t}{2}
                \end{pmatrix}, \\
                &\mathbf{B} =
                    \begin{pmatrix}
                    - i\sqrt{\kappa_a^e} & 0 \\
                    0 & 0 \\
                    0 & 0 \\
                    0 & 0 \\
                    0 & - i\sqrt{\kappa_c^e}
                \end{pmatrix}, \\
                &\mathbf{C} =
                    \begin{pmatrix}
                    - i\sqrt{\kappa_a^e} & 0 & 0 & 0 & 0 \\
                    0 & 0 & 0 & 0 &  - i\sqrt{\kappa_c^e}
                \end{pmatrix}, \\
                &\mathbf{D} =
                    \begin{pmatrix}
                    1 & 0 \\
                    0 & 1
                \end{pmatrix}.
            \end{align}
            Taking the Fourier transform, the input-output relation in the frequency domain is written as $\hat{\mathbf{a}}_{\mathrm{out}}(\omega) = \mathbf{S}(\omega) \hat{\mathbf{a}}_{\mathrm{in}}(\omega)$, where the transfer matrix is $\mathbf{S}(\omega) = \mathbf{C}\left(- i \omega \mathbf{I} - \mathbf{A}\right)^{-1}\mathbf{B} + \mathbf{D}$.
            Here, $\mathbf{I}$ denotes the identity matrix.

        \subsubsection{Conversion efficiency}
            In analogy with the standard cooperativity-based description of quantum transduction \cite{Wang_PRR2022}, we define the frequency-dependent complex cooperativity-like parameters as
            \begin{align}
                C_{ab}(\omega)
                &= g_x^2\chi_a(\omega)\chi_b(\omega), \\
                C_{b1}(\omega)
                &= g_y^2\chi_b(\omega)\chi_1(\omega), \\
                C_{12}(\omega)
                &= |\tilde{g}_{12}|^2\chi_1(\omega)\chi_2(\omega), \\
                C_{2c}(\omega)
                &= g_{02}^2\chi_2(\omega)\chi_c(\omega),
            \end{align}
            where we have introduced the {\it susceptibilities}
            \begin{align}
                \chi_a^{-1}(\omega) &= \kappa_a^t/2 - i(\omega - \omega_{a}), \\
                \chi_b^{-1}(\omega) &= \kappa_b^i/2 - i(\omega - \omega_b), \\
                \chi_1^{-1}(\omega) &= \gamma_{10}/2 - i(\omega - \tilde{\omega}_{10}), \\
                \chi_2^{-1}(\omega) &= (\gamma_{20} + \gamma_{21})/2 - i(\omega - \tilde{\omega}_{20}), \\
                \chi_c^{-1}(\omega) &= \kappa_c^t/2 - i[\omega - (\omega_c - \mu_{\mathrm{p}})].
            \end{align}

            Using these definitions, the microwave-to-optical transduction efficiency is obtained from the corresponding transfer-matrix element $S_{ca}(\omega) = \langle \hat{c}_{\mathrm{out}}(\omega) \rangle / \langle \hat{a}_{\mathrm{in}}(\omega) \rangle$ as
            \begin{align}
                P_{\mathrm{coh}}^{(\mathrm{PM})}(\omega)
                &\equiv \left|S_{ca}(\omega)\right|^2, \nonumber \\
                &= \eta_a(\omega) \eta_c(\omega) \frac{4|C_{ab}(\omega)C_{b1}(\omega)C_{12}(\omega)C_{2c}(\omega)|}{|1 + C_{ab}(\omega) + C_{b1}(\omega) + C_{12}(\omega) + C_{2c}(\omega) + C_{ab}(\omega)C_{12} (\omega)+ C_{ab}(\omega)C_{2c}(\omega) + C_{b1}(\omega)C_{2c}(\omega)|^2},
            \end{align}
            where $\eta_a(\omega) = \kappa_a^e|\chi_a(\omega)|/2$ and $\eta_c(\omega) = \kappa_c^e|\chi_c(\omega)|/2$ are the frequency-dependent coupling efficiencies for the microwave and optical cavities, respectively.

        \subsubsection{Optimal pump amplitude \texorpdfstring{$E_{\mathrm{p}}$}{Ep}}
            For efficient conversion, we employ the quintuply resonant condition $\omega_a = \omega_b = \tilde{\omega}_{10} = \tilde{\omega}_{20} = \omega_c - \mu_{\mathrm{p}}$.
            For an input signal frequency $\mu_{\mathrm{s}}$ resonant with this common frequency, the cooperativity-like parameters become real and positive, which simplifies the optimization of the conversion efficiency.
            Specifically, they reduce to $C_{ab} = 4g_x^2 / (\kappa_a^t \kappa_b^i)$, $C_{b1} = 4g_y^2 / (\kappa_b^i \gamma_{10})$, $C_{12} = 4|\tilde{g}_{12}|^2 / [\gamma_{10}(\gamma_{20} + \gamma_{21})]$ and $C_{2c} = 4g_{02}^2 / [(\gamma_{20} + \gamma_{21})\kappa_c^t]$.
            Neglecting the $E_{\mathrm{p}}$ dependence of $\tilde{\omega}_{10}$ and $\tilde{\omega}_{20}$, the conversion efficiency is maximized by the condition $\frac{\partial}{\partial C_{12}}P_{\mathrm{coh}}^{(\mathrm{PM})}(\mu_{\mathrm{s}}) = 0$.
            The corresponding optimal pump amplitude in the phenomenological model is
            \begin{equation}
                E_{\mathrm{p, opt}} = \sqrt{\frac{(\kappa_c^t/2)^2 + (\mu_{\mathrm{p}} - \omega_c)^2}{\kappa_c^e} \frac{\gamma_{10}(\gamma_{20} + \gamma_{21})}{4g_{12}^2} C_{12}^{\mathrm{opt}}},
            \end{equation}
            where $C_{12}^{\mathrm{opt}} = (1 + C_{2c})(1 + C_{ab} + C_{b1}) / (1 + C_{ab})$.

    \subsection{Numerical results}
    \label{subsec:numerical_results}
        \begin{figure}[htbp]
            \begin{center}
            \includegraphics[width=0.95\columnwidth]{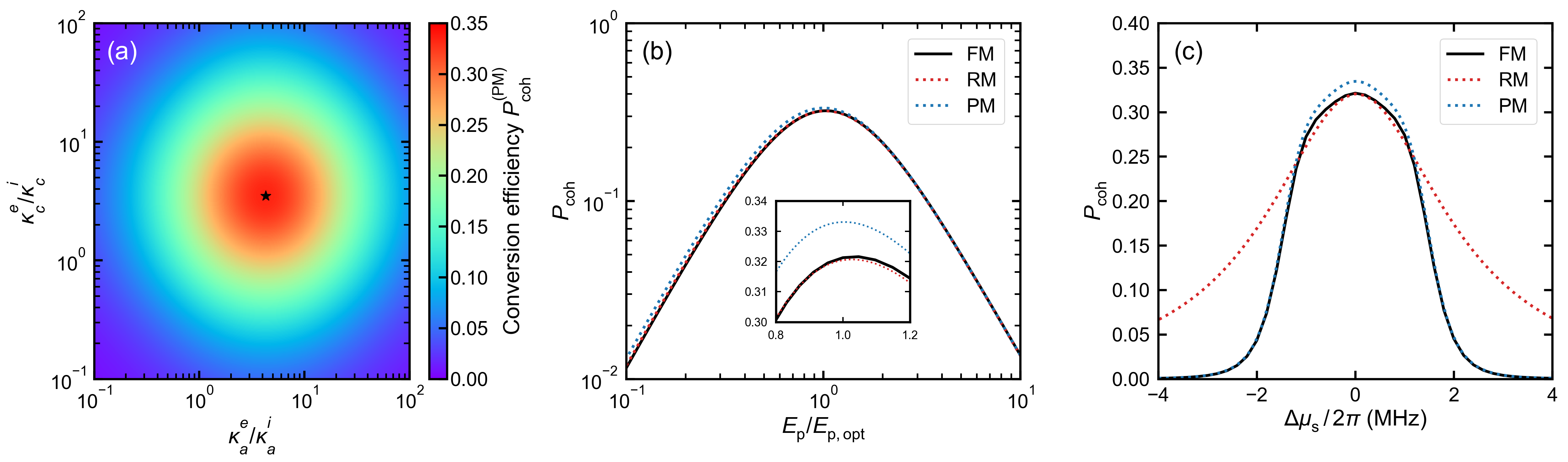}
            \caption{
                Optimization of the external coupling rates $\kappa_a^e$ and $\kappa_c^e$.
                (a) Dependence of the conversion efficiency on $\kappa_a^e$ and $\kappa_c^e$, based on the phenomenological model.
                The black star indicates the optimal point, $(\kappa_a^e/\kappa_a^i,\,\kappa_c^e/\kappa_c^i) = (4.3, 3.5)$.
                (b) Coherent conversion efficiency as a function of the pump amplitude $E_{\mathrm{p}}$ for the full model (FM), reduced model (RM), and phenomenological model (PM).
                The inset shows an enlarged view of the region near the peaks.
                The black solid line and the blue dotted line almost overlap. 
                (c) Coherent conversion efficiency as a function of the input microwave-signal frequency for each model.
                Here, $\Delta\mu_{\mathrm{s}}$ denotes the frequency detuning from resonance, i.e., $\Delta\mu_{\mathrm{s}} = \mu_{\mathrm{s}} - \omega_{10}$. 
            } \label{FIG:numerical_results}
            \end{center}
        \end{figure}
        
        First, we search for the optimal values of external coupling rates $\kappa_a^e$ and $\kappa_c^e$ using the phenomenological model.
        Fig.~\ref{FIG:numerical_results}(a) shows the dependence of the conversion efficiency $P_{\mathrm{coh}}^{(\mathrm{PM})}(\mu_\mathrm{s})$ on $\kappa_a^e$ and $\kappa_c^e$, with the pump amplitude fixed at $E_{\mathrm{p, opt}}$.
        We obtain the optimal values $\kappa_a^e/2\pi = 2.2$~MHz and $\kappa_c^e/2\pi = 180$~GHz, at which the conversion efficiency reaches its maximum value, $P_{\mathrm{coh}}^{(\mathrm{PM})}(\mu_\mathrm{s}) = 0.33$.

        To assess the accuracy of the estimated optimal pump amplitude $E_{\mathrm{p, opt}}$ for given values of $\kappa_a^e$ and $\kappa_c^e$, we calculate the pump-amplitude dependence of coherent conversion efficiencies for the full, reduced and phenomenological models as shown in Fig.~\ref{FIG:numerical_results}(b).
        At the estimated optimal pump amplitude $E_{\mathrm{p, opt}}$, which corresponds to an input power of 48~pW, we obtain the maximum values $P_{\mathrm{coh}}^{\mathrm{(FM)}} = 0.32$, $P_{\mathrm{coh}}^{\mathrm{(RM)}} = 0.32$ and $P_{\mathrm{coh}}^{(\mathrm{PM})}(\mu_\mathrm{s}) = 0.33$ for the full, reduced, and phenomenological models, respectively.
        These results confirm that the three models show good agreement in the (coherent) conversion efficiencies.
        However, the overestimation of the conversion efficiency by the phenomenological model relative to the other two models can be attributed to its neglect of the saturation effect of the color center.

        Next, we determine the conversion bandwidth for each model by calculating the conversion efficiency as a function of the detuning of the input microwave-signal frequency from resonance.
        When the conversion bandwidth is defined as the full width at half maximum of the conversion-efficiency curve in Fig.~\ref{FIG:numerical_results}(c), the bandwidths are found to be $BW^{(\mathrm{FM})} = 3.0$~MHz, $BW^{(\mathrm{RM})} = 5.0$~MHz, and $BW^{(\mathrm{PM})} = 3.0$~MHz.
        The good agreement between $BW^{(\mathrm{FM})}$ and $BW^{(\mathrm{PM})}$, together with the overestimation of $BW^{(\mathrm{RM})}$, can be attributed to the steady-state approximations assumed in the reduced model [see Eqs.~(\ref{eq:ss_approximation_start})-(\ref{eq:ss_approximation_end})], which are justified in the large linewidth limit of the corresponding cavities.

        \begin{figure}[htbp]
            \begin{center}
            \includegraphics[width=0.75\columnwidth]{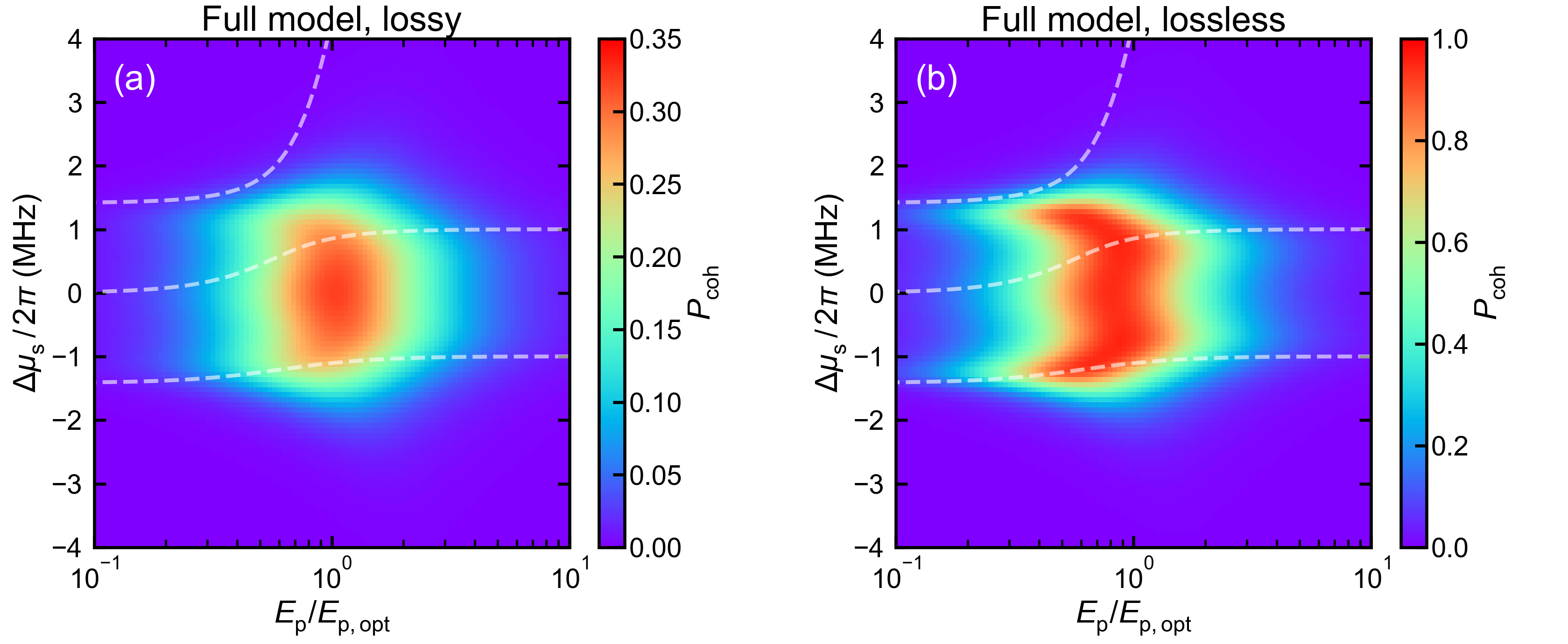}
            \caption{
                Coherent conversion efficiency $P_{\mathrm{coh}}$ as a function of the pump amplitude and input microwave-signal frequency for the full model, calculated (a) with and (b) without internal losses.
                The white dashed lines indicate the eigenvalues of the pump-dressed full-model Hamiltonian in the rotating frame, where the second-, third-, and fourth-lowest eigenvalues are shown.
            } \label{FIG:eigenvalues}
            \end{center}
        \end{figure}

        The bandwidth can also be understood from the eigenstates of the system.
        In Fig.~\ref{FIG:eigenvalues}(a), the conversion efficiency of the transducer consists of three peaks, each of which corresponds to an energy eigenvalue of the system.
        This correspondence becomes clear when internal losses in the resonators and the color center are neglected, as shown in Fig.~\ref{FIG:eigenvalues}(b).
        In the weak-pump regime, the separations between the central peak and the upper and lower peaks approach an approximately constant value of 1.4~MHz, comparable to the microwave-mechanical coupling strength $g_x/2\pi$ and the mechanical-NV$^0$ coupling strength $g_y/2\pi$.

        \begin{figure}[htbp]
            \begin{center}
            \includegraphics[width=0.45\columnwidth]{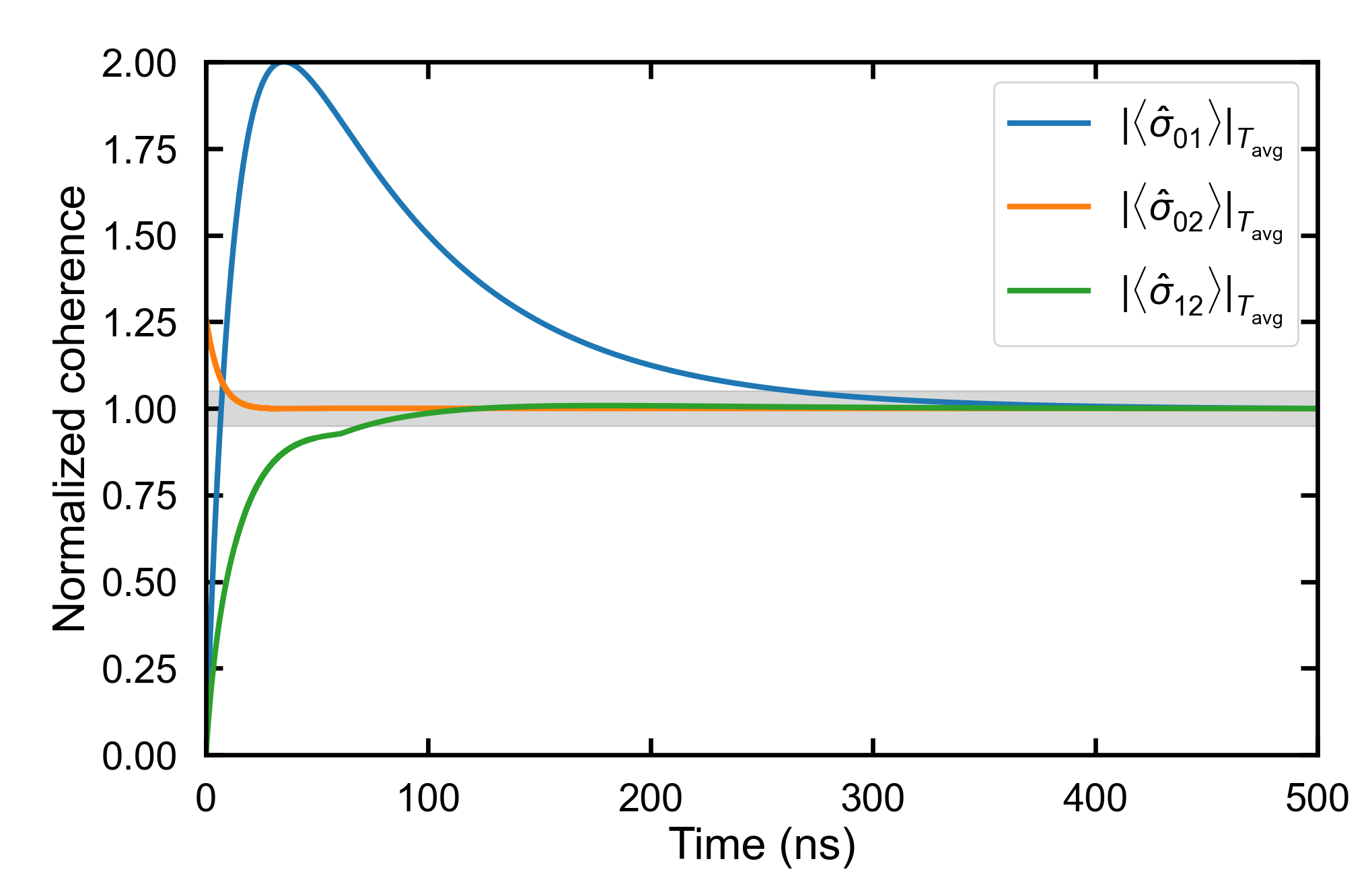}
            \caption{
                Time evolution of the coherence terms of NV$^0$ in the reduced model.
                The gray-shaded region indicates the $\pm 5\%$ range around unity, corresponding to the normalized steady-state value.
                The coherence amplitudes are averaged over a time interval of $T_{\mathrm{avg}} = 100$~ps and normalized by their respective steady-state values.
                The pump amplitude is set to $E_{\mathrm{p,opt}}$.
                The results remain unchanged in the weak-microwave-input regime.
            }
            \label{FIG:time_evolution}
            \end{center}
        \end{figure}

        In this work, we introduce a conversion time $\tau_{\mathrm{conv}}$ to model spectral diffusion and to estimate the dark-count probability.
        To estimate $\tau_{\mathrm{conv}}$, we solve the Heisenberg-Langevin equations of the reduced model, Eqs.~(\ref{eq:reduced_equations_start}--\ref{eq:reduced_equations_end}), in the time domain and calculate the dynamics of the NV$^0$ coherence terms $|\langle\hat{\sigma}_{01}(t)\rangle|$, $|\langle\hat{\sigma}_{02}(t)\rangle|$, and $|\langle\hat{\sigma}_{12}(t)\rangle|$.
        Because the calculated coherence terms exhibit rapid oscillations, we average them over a time interval $T_{\mathrm{avg}}$ according to
        \begin{equation}
            |\langle\hat{\sigma}_{ij}(t)\rangle|_{T_{\mathrm{avg}}} = \frac{1}{T_{\mathrm{avg}}} \int_{t - T_{\mathrm{avg}}/2}^{t + T_{\mathrm{avg}}/2} |\langle\hat{\sigma}_{ij}(t')\rangle|\,dt'.
        \end{equation}

        For simplicity, we consider a weak coherent microwave input.
        As shown in Fig.~\ref{FIG:time_evolution}, the system is regarded as having reached the steady state when all three time-averaged coherence amplitudes remain within $\pm 5\%$ of their respective steady-state values. 
        Under this criterion, the settling time is below approximately $300$~ns.
        We therefore conservatively set $\tau_{\mathrm{conv}} = 300$~ns.

        We further use $\tau_{\mathrm{conv}}$ as the photon-detection time window.
        Assuming that dark counts follow a Poisson process with a rate $R_{\mathrm{d}}$, the dark-count probability per detection window is given by
        \begin{equation}
            P_{\mathrm{d}} = 1-\exp(-\tau_{\mathrm{conv}}R_{\mathrm{d}}) \simeq \tau_{\mathrm{conv}}R_{\mathrm{d}},
        \end{equation}
        where the last approximation is valid for $\tau_{\mathrm{conv}}R_{\mathrm{d}} \ll 1$.
\section{Remote entanglement generation} \label{sec:entanglement_generation}

    \begin{figure}[htbp]
        \centering
        \includegraphics[width=0.5\columnwidth]{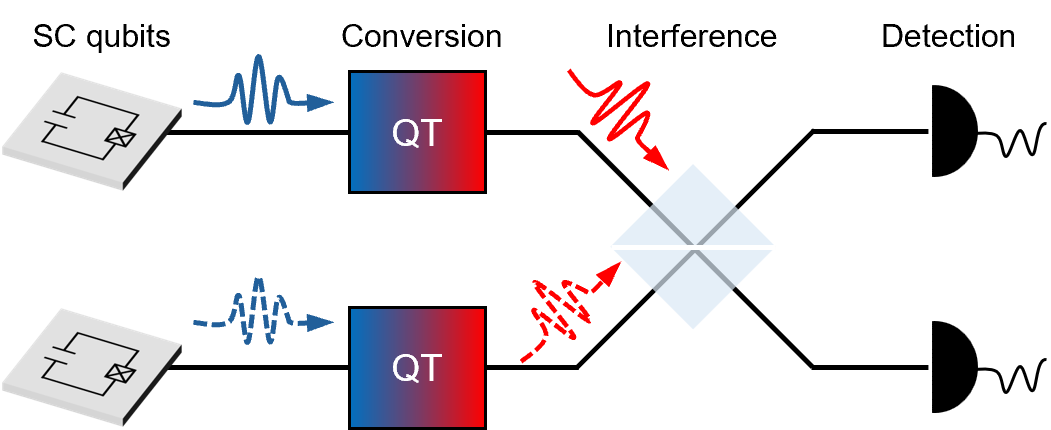}
        \caption{
            Remote entanglement generation. 
            Microwave photons entangled with superconducting qubits (SC qubits) are converted into optical photons by quantum transducers (QTs) and injected into an interferometer, generating remote entanglement. 
        } \label{FIG:remote_entanglement_schematic}
    \end{figure}

    In this work, we consider the generation of entanglement between remote superconducting qubits using the single-click scheme \cite{Cabrillo_PRA1999} as shown in Fig.~\ref{FIG:remote_entanglement_schematic}.
    For symmetric paths, where the two transduction channels have identical efficiencies and noise properties, this protocol has been discussed in Ref.~\cite{Zeuthen_QST2020}.
    However, those results cannot be directly applied to the asymmetric case considered here, in which the two paths may have different conversion efficiencies, dark-count probabilities, and transduction phases.
    In the following, we derive a more general expression that includes path asymmetry, limited conversion coherence and the phase difference between the two paths.

    Generally, in the single-click scheme, an effective Raman-type transition is used for generating qubit-photon entanglement at each node \cite{Kurpiers_Nature2018}.
    Starting from the qubit storage state $|0\rangle_{q}$, a weak excitation creates a photon with probability $P_{\mathrm{e}}$ and transfers the qubit to another long-lived storage state $|1\rangle_{q}$.
    After eliminating the intermediate excited state, the effective qubit-photon state at node $i\,(=\mathrm{A,B})$ is given by $\sqrt{1 - P_{\mathrm{e}}}|0_i\rangle_{q}|0_i\rangle_{p} + \sqrt{P_{\mathrm{e}}} |1_i\rangle_{q}|1_i\rangle_{p}$, where $q$ and $p$ denote the qubit and the photonic modes, respectively, and $|0_i\rangle_{p}$ and $|1_i\rangle_{p}$ represent the zero- and one-photon number states at node $i$, respectively.
    This qubit-photon correlation is essential: after the photonic modes interfere and the state is conditioned on a single detector click, the which-path information is erased and the stationary qubits are projected onto an entangled state.
    When both nodes are excited with the same weak-excitation probability $P_{\mathrm{e}}$, the initial state of the total system is
    \begin{align}
        \label{eq:initial_state}
        \begin{split}
            |\psi_0\rangle     &= (1 - P_{\mathrm{e}})\,|0_{\mathrm{A}}0_{\mathrm{B}}\rangle_{q} |0_{\mathrm{A}}0_{\mathrm{B}}\rangle_{p} + P_{\mathrm{e}} |1_{\mathrm{A}}1_{\mathrm{B}}\rangle_{q} |1_{\mathrm{A}}1_{\mathrm{B}}\rangle_{p} \\
                               &\qquad+ \sqrt{P_{\mathrm{e}}(1 - P_{\mathrm{e}})} |0_{\mathrm{A}}1_{\mathrm{B}}\rangle_{q} |0_{\mathrm{A}}1_{\mathrm{B}}\rangle_{p}
                               + \sqrt{P_{\mathrm{e}}(1 - P_{\mathrm{e}})} |1_{\mathrm{A}}0_{\mathrm{B}}\rangle_{q} |1_{\mathrm{A}}0_{\mathrm{B}}\rangle_{p}.
        \end{split}
    \end{align}

    Next, we model the transduction process at node $i$ as a composite quantum channel $\tilde{\mathcal{E}}_i = \mathcal{N}_i\circ\mathcal{E}_i$.
    Here, $\mathcal{E}_i$ describes photon conversion with total probability $P_{\mathrm{tot}}^{(i)}$ and an associated phase rotation $\theta_i$, whereas $\mathcal{N}_i$ describes the addition of a noise photon with probability $P_{\mathrm{d}}^{(i)}$.
    The conversion process is decomposed into coherent and incoherent parts: $\mathcal{E}_i$ performs coherent conversion with probability $P_{\mathrm{coh}}^{(i)}$ and incoherent conversion with probability $P_{\mathrm{inc}}^{(i)}$.
    Thus, the possible transducer outputs are a coherent photon in mode $p_i$, an incoherent photon in mode $p'_i$, and a noise photon in mode $d_i$.
    The probability conservation condition $P_{\mathrm{coh}}^{(i)} + P_{\mathrm{inc}}^{(i)} = P_{\mathrm{tot}}^{(i)}$ reflects that, for a single input photon, coherent conversion into $p_i$ and incoherent conversion into $p'_i$ are alternative outcomes.

    With this notation, the conversion channel $\mathcal{E}_i$ is represented, on the relevant input subspace where $p'_i$ is initially in the vacuum state, by the following Kraus operators:
    \begin{align}
        \hat{K}_i^{(0)} &= |0_i\rangle_p |0_i\rangle_{p'}\, {}_p\langle 0_i|\, {}_{p'}\langle 0_i| + \sqrt{P_{\mathrm{coh}}^{(i)}}\,e^{i\theta_i}\,|1_i\rangle_p |0_i\rangle_{p'}\, {}_p\langle 1_i|\, {}_{p'}\langle 0_i|,
        \\
        \hat{K}_i^{(1)} &= \sqrt{P_{\mathrm{inc}}^{(i)}}\,e^{i\theta_i}\,|0_i\rangle_p |1_i\rangle_{p'}\, {}_p\langle 1_i|\, {}_{p'}\langle 0_i|,
        \\
        \hat{K}_i^{(2)} &= \sqrt{1-P_{\mathrm{tot}}^{(i)}}\,|0_i\rangle_p |0_i\rangle_{p'}\, {}_p\langle 1_i|\, {}_{p'}\langle 0_i|.
    \end{align}
    The noise-addition channel $\mathcal{N}_i$ is represented by
    \begin{align}
        \hat{N}_{i}^{(0)} &= \sqrt{1 - P_{\mathrm{d}}^{(i)}}\,|0_i\rangle_{d}\,{}_{d}\langle0_i| + |1_i\rangle_{d}\,{}_{d}\langle1_i|, \\
        \hat{N}_{i}^{(1)} &= \sqrt{P_{\mathrm{d}}^{(i)}}\,|1_i\rangle_{d}\,{}_{d}\langle0_i|,
    \end{align}
    where we assume that the noise mode $d_i$ is initially in the vacuum state and restrict it to the zero- and one-photon subspace.
    Therefore, the Kraus operators of the composite transducer channel $\tilde{\mathcal{E}}_i = \mathcal{N}_i\circ\mathcal{E}_i$ are
    \begin{equation}
        \tilde{K}_{i}^{(m,k)} = \hat{N}_{i}^{(k)} \hat{K}_{i}^{(m)} \quad (m = 0,1,2;\quad k = 0,1).
    \end{equation}

    After transduction, the converted photons from the two nodes are combined at a 50:50 beam splitter, which is represented by an isometry $\hat{W}_{\mathrm{BS}}$.
    We assume that only the coherent photons in the modes $p_{\mathrm{A}}$ and $p_{\mathrm{B}}$ are indistinguishable and interfere at the beam splitter.
    By contrast, the incoherent photons and noise photons associated with different nodes are assumed to be distinguishable and therefore do not interfere with each other.
    Under this assumption, the beam-splitter isometry can be decomposed as $\hat{W}_{\mathrm{BS}} = \hat{W}_{p} \otimes \hat{W}_{p'_\mathrm{A}} \otimes \hat{W}_{p'_\mathrm{B}} \otimes \hat{W}_{d_\mathrm{A}} \otimes \hat{W}_{d_\mathrm{B}}$.
    Let $\mathrm{C}$ and $\mathrm{D}$ denote the two output ports of the beam splitter.
    On the subspaces that can be occupied before and after the beam splitter, the corresponding isometries are given by
    \begin{align}
        \hat{W}_p               &= |0_{\mathrm{C}}0_{\mathrm{D}}\rangle_{p}\,{}_{p}\langle0_{\mathrm{A}}0_{\mathrm{B}}| + \frac{1}{\sqrt2} \Big( |2_{\mathrm{C}}0_{\mathrm{D}}\rangle_{p}\,{}_{p}\langle1_{\mathrm{A}}1_{\mathrm{B}}| - |0_{\mathrm{C}}2_{\mathrm{D}}\rangle_{p}\,{}_{p}\langle1_{\mathrm{A}}1_{\mathrm{B}}|\Big ) \nonumber \\
                                &\qquad + \frac{1}{\sqrt2}\Big( |1_{\mathrm{C}}0_{\mathrm{D}}\rangle_{p}\,{}_{p}\langle1_{\mathrm{A}}0_{\mathrm{B}}|+|1_{\mathrm{C}}0_{\mathrm{D}}\rangle_{p}\,{}_{p}\langle0_{\mathrm{A}}1_{\mathrm{B}}|+|0_{\mathrm{C}}1_{\mathrm{D}}\rangle_{p}\,{}_{p}\langle1_{\mathrm{A}}0_{\mathrm{B}}|-|0_{\mathrm{C}}1_{\mathrm{D}}\rangle_{p}\,{}_{p}\langle0_{\mathrm{A}}1_{\mathrm{B}}| \Big), \\
        \hat{W}_{X_\mathrm{A}} &= |0_{\mathrm{C}}0_{\mathrm{D}}\rangle_{X_\mathrm{A}}\,{}_{X_\mathrm{A}}\langle0_{\mathrm{A}}0_{\mathrm{B}}| + \frac{1}{\sqrt2} \Big( |1_{\mathrm{C}}0_{\mathrm{D}}\rangle_{X_\mathrm{A}}\,{}_{X_\mathrm{A}}\langle1_{\mathrm{A}}0_{\mathrm{B}}| + |0_{\mathrm{C}}1_{\mathrm{D}}\rangle_{X_\mathrm{A}}\,{}_{X_\mathrm{A}}\langle1_{\mathrm{A}}0_{\mathrm{B}}| \Big), \\
        \hat{W}_{X_\mathrm{B}} &= |0_{\mathrm{C}}0_{\mathrm{D}}\rangle_{X_\mathrm{B}}\,{}_{X_\mathrm{B}}\langle0_{\mathrm{A}}0_{\mathrm{B}}| + \frac{1}{\sqrt2} \Big( |1_{\mathrm{C}}0_{\mathrm{D}}\rangle_{X_\mathrm{B}}\,{}_{X_\mathrm{B}}\langle0_{\mathrm{A}}1_{\mathrm{B}}| - |0_{\mathrm{C}}1_{\mathrm{D}}\rangle_{X_\mathrm{B}}\,{}_{X_\mathrm{B}}\langle0_{\mathrm{A}}1_{\mathrm{B}}| \Big),
    \end{align}
    where $X_{\mathrm{A}}\in\{p'_{\mathrm{A}},d_{\mathrm{A}}\}$ and $X_{\mathrm{B}}\in\{p'_{\mathrm{B}},d_{\mathrm{B}}\}$.

    For the coherent mode $p$, two photons can simultaneously enter the beam splitter from the two input ports.
    Since these photons are indistinguishable, they undergo Hong--Ou--Mandel bunching \cite{Hong_PRL1987}, as described by the transformation of $|1_{\mathrm A}1_{\mathrm B}\rangle_{p}$ in $\hat{W}_p$.
    Such two-photon interference does not occur for the other modes $X_{\mathrm{A}}$ and $X_{\mathrm{B}}$, because photons in these modes are treated as distinguishable.

    For photon detection, we assume that the photon detectors at the output ports $\mathrm{C}$ and $\mathrm{D}$ perform threshold detection.
    Namely, in each output port, the detector gives a no-click outcome if the total photon number is zero, and a click outcome if the total photon number is one or larger.
    The detector does not distinguish among the photonic modes $(p,p',d)$.

    Since the measurement operator corresponding to the no-click outcome at output port $i\,(=\mathrm{C},\mathrm{D})$ is $\hat{\Pi}_i^{(0)} = |0_i\rangle_{p}\langle0_i| \otimes |0_i\rangle_{p'}\langle0_i| \otimes |0_i\rangle_{d}\langle0_i|$, the measurement operator corresponding to the click outcome is $\hat{\Pi}_i^{(\geq 1)} = \hat{I}_i - \hat{\Pi}_i^{(0)}$.
    Here, $\hat{I}_i$ denotes the identity operator on the joint Hilbert space of the modes $(p_i,p'_i,d_i)$ at output port $i$.
    The measurement operators corresponding to a click in one output port and no click in the other are then given by $\hat{\Pi}_{\mathrm{C}} = \hat{\Pi}_{\mathrm{C}}^{(\geq 1)} \otimes \hat{\Pi}_{\mathrm{D}}^{(0)}$ and $\hat{\Pi}_{\mathrm{D}} = \hat{\Pi}_{\mathrm{C}}^{(0)} \otimes \hat{\Pi}_{\mathrm{D}}^{(\geq 1)}$.
    Hence, the POVM element corresponding to the single-click event, in which exactly one of the two detectors clicks, is
    \begin{align}
        \hat{\Pi}_{\mathrm{1c}} &= \hat{\Pi}_{\mathrm{C}} + \hat{\Pi}_{\mathrm{D}}.
    \end{align}

    Let $\hat{\rho}_{0} = |\psi_0\rangle\langle\psi_0|$ be the density matrix of the initial state in Eq.~(\ref{eq:initial_state}).
    The density matrix of the qubits conditioned on the event that exactly one of the two detectors clicks is then given by
    \begin{equation}
        \hat{\rho}_{\mathrm{1c}} = \frac{\sum_{m,n,k,l} \mathrm{Tr}_{p,p',d}\left[ \hat{\rho}_{0}\,(\tilde{K}^{(mn,kl)})^{\dagger}\,(\hat{W}_{\mathrm{BS}})^{\dagger}\,\hat{\Pi}_{\mathrm{1c}}\,(\hat{W}_{\mathrm{BS}})\,(\tilde{K}^{(mn,kl)}) \right]}{\sum_{m,n,k,l} \mathrm{Tr}\left[ \hat{\rho}_{0}\,(\tilde{K}^{(mn,kl)})^{\dagger}\,(\hat{W}_{\mathrm{BS}})^{\dagger}\,\hat{\Pi}_{\mathrm{1c}}\,(\hat{W}_{\mathrm{BS}})\,(\tilde{K}^{(mn,kl)}) \right]}.
    \end{equation}
    Here, $\mathrm{Tr}_{p,p',d}$ denotes the partial trace over all photonic modes $p$, $p'$, and $d$.
    $\tilde{K}^{(mn,kl)} \equiv \tilde{K}^{(m,k)}_{\mathrm{A}} \otimes \tilde{K}^{(n,l)}_{\mathrm{B}}$ denotes the two-path Kraus operator defined on the tensor-product Hilbert space.

    When the photons emitted from the two paths are indistinguishable, the which-path information is erased, and photon detection projects the qubits onto an entangled state.
    For the initial state $|\psi_0\rangle$, the ideal entangled states generated by the single-click event are $|\Psi^{\pm}\rangle_{q} = (|1_{\mathrm{A}}0_{\mathrm{B}}\rangle_{q} \pm |0_{\mathrm{A}}1_{\mathrm{B}}\rangle_{q})/\sqrt{2}$, where the relative phase is determined by which detector clicks.
    If a phase flip is applied when the detector at port $\mathrm{D}$ clicks, the target Bell state can be chosen as $|\Psi^{+}\rangle_{q}$.
    After applying this detector-dependent phase correction, we denote the resulting conditional qubit state by $\hat{\rho}'_{\mathrm{1c}}$ and define the fidelity as $F_{\mathrm{1c}} \equiv \langle\Psi^{+}|\hat{\rho}'_{\mathrm{1c}}|\Psi^{+}\rangle_{q}$.
    The single-click fidelity is then obtained as
    \begin{align}
        F_{\mathrm{1c}} &= \frac{1}{N} \Big[ \Pi_{10} + \Pi_{01} +  \alpha\,\sqrt{ P_{\mathrm{coh}}^{(\mathrm{A})}P_{\mathrm{coh}}^{(\mathrm{B})} } \cos\Delta\theta\, \Big]\,P_{\mathrm{e}}(1 - P_{\mathrm{e}}), \\
        N               &= 2(1 - P_{\mathrm{e}})^2\,\Pi_{00} + 2P_{\mathrm{e}}(1 - P_{\mathrm{e}})\,\Pi_{10} + 2P_{\mathrm{e}}(1 - P_{\mathrm{e}})\,\Pi_{01} + 2P_{\mathrm{e}}^2\,\Pi_{11}.
    \end{align}
    Here,
    \begin{align}
        \Pi_{00} &= \alpha - \beta, \\
        \Pi_{10} &= \alpha\left( 1 - \frac{P_{\mathrm{tot}}^{(\mathrm{A})}}{2} \right) - (1 - P_{\mathrm{tot}}^{(\mathrm{A})})\beta, \\
        \Pi_{01} &= \alpha\left( 1 - \frac{P_{\mathrm{tot}}^{(\mathrm{B})}}{2} \right) - (1 - P_{\mathrm{tot}}^{(\mathrm{B})})\beta, \\
        \Pi_{11} &= \alpha\gamma + (1 - P_{\mathrm{tot}}^{(\mathrm{A})})(1 - P_{\mathrm{tot}}^{(\mathrm{B})})(\alpha - \beta),
    \end{align}
    where
    \begin{align}
        \alpha &= \left( 1 - \frac{P_{\mathrm{d}}^{(\mathrm{A})}}{2} \right)\left( 1 - \frac{P_{\mathrm{d}}^{(\mathrm{B})}}{2} \right), \\
        \beta  &= \left( 1 - P_{\mathrm{d}}^{(\mathrm{A})} \right)\left( 1 - P_{\mathrm{d}}^{(\mathrm{B})} \right), \\
        \gamma &= \frac{1}{2} \Big[ 1 - (1 - P_{\mathrm{tot}}^{(\mathrm{A})})(1 - P_{\mathrm{tot}}^{(\mathrm{B})}) \Big] - \frac{1}{4} \Big[ P_{\mathrm{tot}}^{(\mathrm{A})}P_{\mathrm{tot}}^{(\mathrm{B})}  - P_{\mathrm{coh}}^{(\mathrm{A})}P_{\mathrm{coh}}^{(\mathrm{B})} \Big].
    \end{align}
    Here, $\Delta\theta$ is the phase difference between the two interference paths induced by the transducers, as defined in Eq.~(\ref{eq:phase_difference}).
    Since we consider the entanglement generation over short distances of about 10~m for distributed quantum computing, we neglect losses other than those due to the transducers and assume that the interference paths themselves are phase locked.
    For fixed transduction and noise parameters, the fidelity $F_{\mathrm{1c}}$ is maximized when $P_{\mathrm{e}}$ is given by
    \begin{equation}
        P_{\mathrm{e}} = \frac{\sqrt{\Pi_{00}}}{\sqrt{\Pi_{00}} + \sqrt{\Pi_{11}}}.
    \end{equation}
    When the dark-count probabilities $P_{\mathrm{d}}^{(\mathrm{A})}$ and $P_{\mathrm{d}}^{(\mathrm{B})}$ are negligible, we obtain
    \begin{equation}
        F_{\mathrm{1c}} = \frac{1}{2} + \frac{\sqrt{ P_{\mathrm{coh}}^{(\mathrm{A})}P_{\mathrm{coh}}^{(\mathrm{B})} }}{P_{\mathrm{tot}}^{(\mathrm{A})} + P_{\mathrm{tot}}^{(\mathrm{B})}} \cos \Delta\theta.
    \end{equation}

    Since $N$ denotes the total single-click probability, the heralding rate is approximately given by
    \begin{equation}
        R_{\mathrm{herald}} \sim N R_{\mathrm{rep}},
    \end{equation}
    where $R_{\mathrm{rep}}$ is the repetition rate of the entanglement generation attempt.

\end{document}